\documentclass[2col]{aa}

\usepackage{graphicx}
\usepackage{amsmath}
\usepackage{amsfonts}
\usepackage{wasysym}
\usepackage{comment}
\usepackage{booktabs}

\title{Rainy downdrafts in abyssal atmospheres}
\author{S. Markham
\inst{1}
\and
T. Guillot
\inst{1}
\and
C. Li
\inst{2}}

\institute{Universit\'{e} C\^{o}te d'Azur, Observatoire de la C\^{o}te d'Azur, CNRS, Laboratoire Lagrange, Nice, France\\
\email{steve.markham@oca.eu}
\and
University of Michigan, Dept. of Climate and Space Science Engineering, Ann Arbor, MI, USA}
\date{\today}

\begin{document}

\abstract 
{Results from Juno's microwave radiometer indicate non-uniform mixing of ammonia vapor in Jupiter's atmosphere down to tens of bars, far beneath the cloud level. 
Helioseismic observations suggest solar convection may require narrow, concentrated downdrafts called entropy rain to accommodate the Sun's luminosity. 
Both observations suggest some mechanism of non-local convective transport.
}
{We seek to predict the depth that a concentrated density anomaly can reach before efficiently mixing with its environment in bottomless atmospheres.}
{We modify classic self-similar analytical models of entraining thermals to account for the compressibility of an abyssal atmosphere. 
We compare these models to the output of high resolution three dimensional fluid dynamical simulations to more accurately model the chaotic influence of turbulence.}
{We find that localized density anomalies propagate down to $\sim 3-8$ times their initial size without substantially mixing with their environment. 
Our analytic model accurately predicts the initial flow, but the self-similarity assumption breaks down after the flow becomes unstable at a characteristic penetration depth. 
}
{In the context of Jupiter, our findings suggest that precipitation concentrated into localized downdrafts of size $\sim$20~km can coherently penetrate to on the order of a hundred kilometers (tens of bars) beneath its initial vaporization level without mixing with its environment. 
This finding is consistent with expected convective storm length-scales, and Juno MWR measurements of ammonia depletion. 
In the context of the Sun, we find that turbulent downdrafts in abyssal atmospheres cannot maintain their coherence through the Sun's convective layer, a potential challenge for the entropy rain hypothesis.
}

\maketitle

\section{Introduction}
\label{sec:intro}

Mixing-length theory (MLT) for turbulent convection predicts efficient stirring by space-filling homogenous convection \citep[e.g.,][]{chan-sofia1987}. 
On this basis, a common assumption in astrophysics and planetary science has been that convective layers in stars and in fluid planets should be well-mixed. 
Condensing species were expected to become uniformly mixed rapidly at depths where they are entirely in vapor form. 
However,  in Jupiter's deep atmosphere, ammonia appears to be generally depleted and variable in abundance down to at least four scale heights below its condensation level \citep{li+2017, depater+2019, bolton+2021}. 
Water also has a variable abundance much deeper than its cloud base, as shown by Galileo probe measurements \citep{wong+2004}, infrared spectroscopy \citep{bjoraker+2015, bjoraker+2018}, and indirect measurements from the Juno Microwave Radiometer (MWR) \citep{li+2020}. 
In the Sun, the convective flow velocities measured by helioseismology appear to be poorly described by MLT \citep{hanasoge+2012}. 
The possibility that localized downdrafts may maintain their coherence without mixing over significant distances has thus been proposed in both the  Jovian \citep{guillot+2020ii} and solar \citep{anders+2019} cases: 
In Jupiter, storms driven by the condensation of water have long been known to be a powerful source of energy, leading to spatially localized intermittent events that transport large amounts of energy in a short time \citep[e.g.,][]{gierasch+2000}. 
Storms yield precipitation in the form of rain, ice, or for Jupiter, ammonia-water hail (``mushballs''). 
Upon evaporation, precipitation leads to a localized increase in mean-molecular weight and decrease in temperature, both effects strongly favoring the formation of a localized downdraft \citep{guillot+2020i, guillot+2020ii}. \\

In the Sun, there is a long history of modeling the effect of compressibility on convection \citep[e.g.,][]{hurlburt+1984, brummell+2002}, focused primarily on large-scale flows.
However, more recent helioseismic measurements indicate  flow velocities that are orders-of-magnitude smaller than would be expected from MLT in order to be compatible with the Sun's luminosity \citep{hanasoge+2012}.
It has been suggested that a modification to MLT that includes an additional non-local mixing term called ``entropy rain'' can reconcile these observations to theory \citep{brandenburg2016}. 
Entropy rain refers to localized entropy anomalies that can sink through long distances, even through the whole  convective region of the Sun, without mixing with their surroundings as MLT would ordinarily posit. 
Therefore the total convective heat flux can be greater than the component of heat flux transported by the large-scale flow field to which helioseismic measurements are sensitive. 
Such a mechanism could reconcile the \cite{hanasoge+2012} observations with the Sun's observed luminosity. 
Simulations of laminar downwelling thermals in a highly compressible environment find that thermals form coherent vortex rings that maintain their concentration and do not mix with their environment \citep{anders+2019}, offering a plausible mechanism for entropy rain. 
This phenomenon motivates more careful analysis of the effect of compressibility on small-scale downdrafts beneath the resolution of larger-scale simulations.
We will therefore compare our findings here, which consider the influence of turbulence, to this prior work that considers the laminar case. 
 \\

Global circulation models of Jupiter including sub-scale moist convection parametrized from the best available models of the Earth's atmosphere have led to abundance variations that are modest and very rapidly negligeable below the water condensation level \citep{delgenio-mcgrattan1990}. 
But giant planets bear at least two important differences from Earth: the absence of a surface, and the low molecular weight of the dominant gas species. 
The absence of a surface forces us to seriously grapple with the chaotic dynamics of moist convection operating over multiple scale heights, a situation that never arises on the Earth. 
Furthermore the low molecular weights of hydrogen and helium imbue condensing species with an additional forcing mechanism, weighing down wet updrafts \citep[see][]{guillot1995} and imbuing wet downdrafts with greater power. 
Also, while the physics of compressible convection has been investigated in some detail in the solar case \citep[e.g.,][]{hurlburt+1984, brummell+2002, nordlund+2009}, the influence of intermittency has not received sufficient attention. 
We therefore seek to model mesoscale convection from isolated sources in abyssal adiabatic atmospheres, to determine whether we should expect the rainfall from a concentrated, energetic storm to efficiently mix, or whether the rainfall can remain coherent and unmixed for some distance. \\

The topic is important: Our Solar System has four planets with abyssal atmospheres, planets with substantial gas envelopes probably constitute the majority of planets in the Galaxy \citep[e.g.,][]{zhu-dong2021}, and most of them have clouds \citep[e.g.,][]{helling2019, loftus+2019}. 
Furthermore, the physics we investigate here should also apply to the atmospheres of brown dwarfs \citep[e.g.,][]{helling-casewell2014}. 
Storm activity is also routinely observed in all of the solar system's giant planets: Thunderstorms have been extensively observed \citep[e.g.,][]{borucki+1982, little+1999, kolmasova+2018, becker+2020} on Jupiter. 
Saturn has an extreme case of decadally intermittent storm activity in its Great White Spot \citep{heath-mckim1990, fischer+2011, li-ingersoll2015}, and large methane storms have been observed from Earth on Uranus \citep{depater+2015} and Neptune \citep{molter+2017}. 
The storms we can see from visible clouds may only scratch the surface; abyssal rock storms on Jupiter could be responsible for its seismicity \citep{markham-stevenson2018}, and lightning strikes inferred from Voyager's plasma wave instrument \citep{gibbard+1999} suggest moist convection may be active at the water cloud layer on Neptune, hundreds of kilometers beneath the visible surface. 
Such ubiquity entreats deeper theoretical understanding of abyssal convection, which appears to be key to properly contextualizing the relationship between stellar and giant planet atmospheres and their deep interiors. \\

To attack this problem, we begin by introducing some simple intuitive models for negatively buoyant thermals in Sec.~\ref{sec:turner}.  that we can use as a heuristic basis, modifying existing self-similar entrainment theories from \cite{turner-book} to account for the tremendous compressibility of abyssal atmospheres. 
Armed with some basic intuition, we then resort to three dimensional fluid dynamical simulations in Sect.~\ref{sec:snap} from Simulating Non-hydrostatic Atmospheres in Planets (SNAP) \citep{snap} in order to account for the effect of self-induced turbulence of strongly negatively buoyant downdrafts. 
In Sec.~\ref{sec:results} we summarize our results, including computational performance and non-dimensional scaling relationships that we apply to particular celestial objects in Sect.~\ref{sec:applications}. 
Then in Sec.~\ref{sec:confounding} we address some of the deficiencies of our simplified model assumption, for example to inspect the importance of our choice of initial conditions, the effect of environmental turbulence, and vertical wind shear. 
We constrain the degree to which these complications affect our conclusions. 
Finally in Sec.~\ref{sec:discussion} we discuss our results in the context of Jupiter and beyond, motivate further observations of Uranus, and chart a course for future steps to more fully understand moist convection in abyssal atmospheres.

\section{Analytical model}
\label{sec:turner}
\begin{figure}
\centering
\includegraphics[scale=.5]{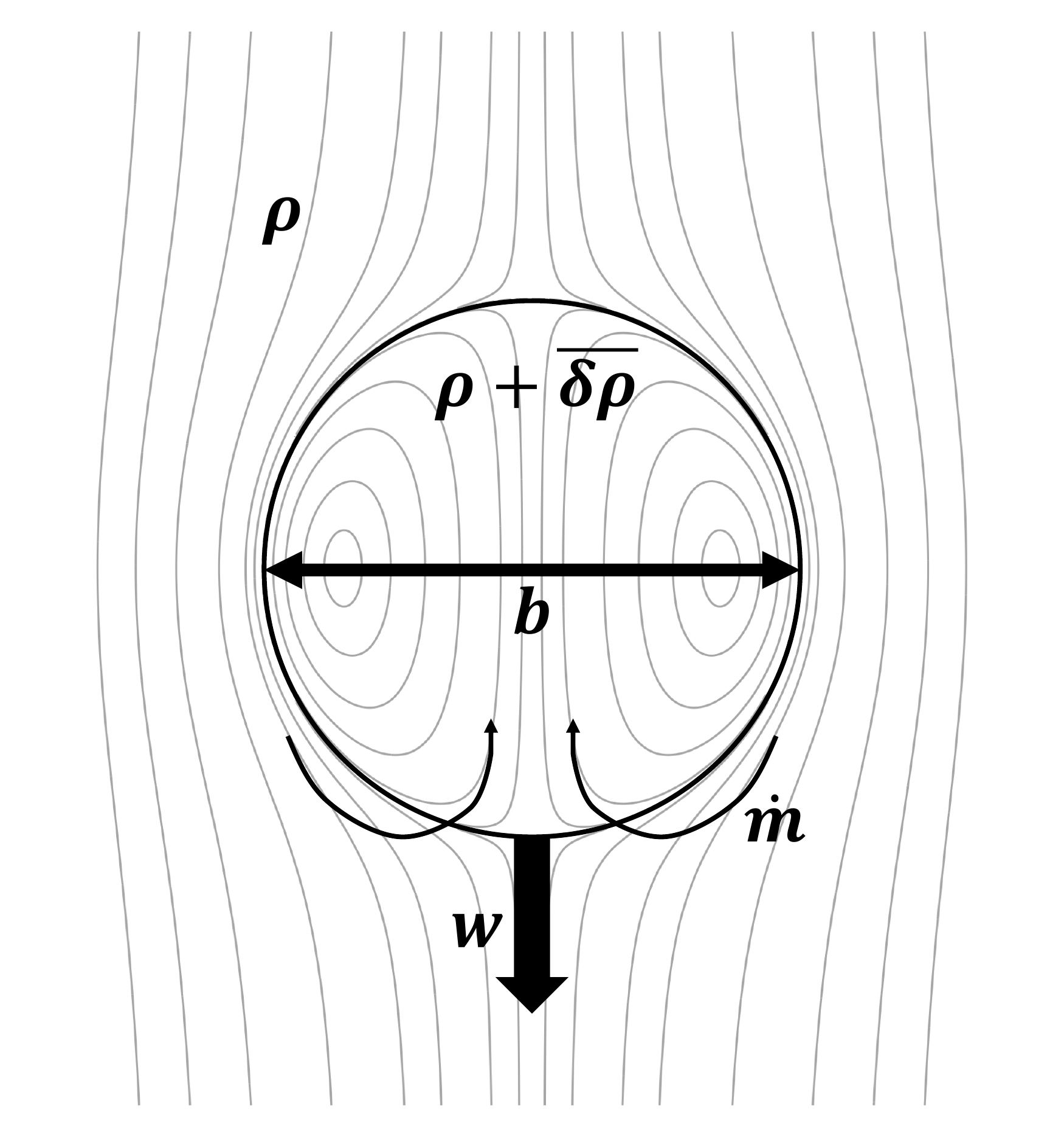}
\caption{Diagram showing the important quantities in our problem overlaid on the stream function of a Hill's spherical vortex from Eq.~\ref{eq:hills-vortex}. 
A spherical vortex ring of diameter $b$ falls at speed $w$ through a medium of density $\rho$.  Within the vortex, there is a mean density perturbation of $\delta \rho$, with $g' \equiv g \overline{\delta \rho} / (\rho + \overline{\delta \rho}) $. 
The vortex entrains external fluid at a rate $\dot{m} = \frac{\pi}{2} b^2 \alpha \rho w$.}
\label{fig:diagram}
\end{figure}

Our analytical model will closely follow the classic work \cite{turner-book}, which deftly addresses many of the main problems of the fluid dynamics of buoyancy from isolated sources. 
This framework provides not only a convenient heuristic, but also experimental corroboration and determination of free parameters that would be prohibitive to constrain on the basis of self-similar theory alone. 
We modify Turner's self-similar equations to account for the compressibility of abyssal atmospheres, and use these results as a starting point to compare against more detailed, physics-based simulations. 
We will model rainy downdrafts as downward-propagating thermals. 
Thermals have been observed to be a general feature of convection, constituting the smallest unit of self-organized convection. 
Following \cite{turner1963}, we characterize the thermal as an entraining vortex ring with a definite size. 
A vortex ring is a flow structure characterized by a mean propagation velocity and circulation around an azimuthally symmetric, dynamically stable ring. 
The simplest analytical model for a vortex ring is Hill's spherical vortex, whose stream function obeys 
\begin{equation}
\psi = \left\{
	\begin{array}{ll}
		-\frac{3w}{4} \left(1 - \frac{4r^2}{b^2} \right) r^2 \sin^2\theta & \quad r \leq b/2 \\
		\frac{w}{2} \left(1 - \frac{b^3}{8r^2} \right) r^2 \sin^2 \theta & \quad r \geq b/2
	\end{array}
\right.
\label{eq:hills-vortex}
\end{equation}
where $w$ is the propagation velocity and $b$ is the vortex ring diameter. 
Thermals have been shown to propagate as vortex rings in laboratory studies \citep[e.g.,][]{turner1957, shusser-gharib2000, gao-yu2016}, and cumulus storms on Earth likewise appear to resemble vortex rings \citep{wang-geerts2013}. 
Figure~\ref{fig:diagram} shows a schematic of the model, with streamlines of Hill's spherical vortex shown in gray lines with important parameters labeled. 
We assume the thermal should remain self-similar as it propagates \citep{turner1957}. 
The central assumption of the analytical model is that the mean rate of inflow into propagating vortex ring should be proportional to its propagation velocity according to some dimensionless entrainment coefficient $\alpha$. \\

We now seek to model the evolution of the thermal as it propagates under these assumptions, with the additional assumption that the environment is of uniform density, and the Boussinesq approximation (we will relax these assumptions later). 
Under our assumptions the entrainment rate is 
\begin{equation}
\frac{dm}{dt} = \frac{\pi}{2} b^2 \alpha  \rho w,
\label{eq:entrainment}
\end{equation}
where $w$ is the propagation velocity, $b$ is the diameter of the vortex ring, and $m$ is the mass enclosed within the Hill's spherical vortex. 
The entrainment coefficient has been empirically determined to match laboratory results for thermals when $\alpha = 1/4$ \citep{turner-book}, but it can take other values in different physical systems and in general is an unconstrained scaling coefficient of order unity that must be empirically determined. 
Under these assumptions, we can write the evolution equation for the vortex ring size as it propagates, 
\begin{equation}
\frac{db^3}{dt} = 3 \alpha b^2 w.
\label{eq:size-incompressible}
\end{equation}
From Equation~\ref{eq:size-incompressible}, we can immediately see that $b \propto \alpha z$, where $w = dz/dt$. 
In other words, $db/dz = \alpha$. 
So according to this model, in an incompressible medium the thermal diameter increases linearly with propagation distance.
Thus $\alpha$ is equivalent to the radian half-angle spread of the cone that encapsulates the entraining spherical vortex as it propagates. \\

Now we can solve for the equation of motion using Newton's second law for momentum, so that 
$
\frac{d}{dt}[m w] = g \overline{\delta \rho} \pi b^3 / 6, 
$
where $m = (\rho + \overline{\delta \rho}) \pi b^3 / 6$ is the mass enclosed within the thermal. 
Then under the Boussinesq approximation, 
\begin{equation}
\frac{d(b^3 w)}{dt} = b^3 g'.
\label{eq:velocity-incompressible}
\end{equation}
where $g' \equiv \frac{\delta \rho}{\rho_1} g$. 
In this model we assume gravity $g$ to be constant, although $g'$ can vary as the density perturbation in the thermal varies. \\

Finally we solve for evolution of the buoyancy, assuming entrainment according to Eq.~\ref{eq:entrainment} and zero detrainment. 
$\frac{dg'}{dt} = \frac{g}{\rho} \frac{d(\Delta \rho)}{dt} = \frac{g}{\rho} \frac{d \rho_1}{dm} \frac{dm}{dt}$ where $\rho_1 \equiv \rho + \Delta \rho$. 
Then if the mass in the thermal is $m\equiv \rho_1 V$, $m + dm = (\rho_1 + d\rho_1) (V + dV) = (\rho_1 + d \rho_1)(V + dm/\rho)$ if the pressure is the same inside and outside of the thermal. 
Solving for $d\rho_1$ to first order in $dm$, we find $\frac{dg'}{dt} = - \frac{3 \alpha}{b} \frac{\rho_1}{\rho} g' w$ from Eq.~\ref{eq:entrainment}. 
Then under the Boussinesq approximation $b^3 \frac{dg'}{dt} = - g' \frac{db^3}{dt}$ from Equation~\ref{eq:size-incompressible}. 
Therefore from the chain rule 
\begin{equation}
\frac{d(b^3 g')}{dt} = 0.
\label{eq:buoyancy-incompressible}
\end{equation}
Intuitively, this means any volume increases due to entraining surrounding material are compensated by a corresponding decrease in $\Delta \rho$ so that the integrated total buoyancy anomaly is conserved. 
This result applies to the neutrally stratified medium. 
In general for a thermal propagating through a stratified medium, one must account for the Br\"{u}nt-V\"{a}is\"{a}l\"{a} frequency of the background material, discussed in Section~\ref{subsec:stratification}. \\

\begin{figure}
\centering
\includegraphics[scale=.29]{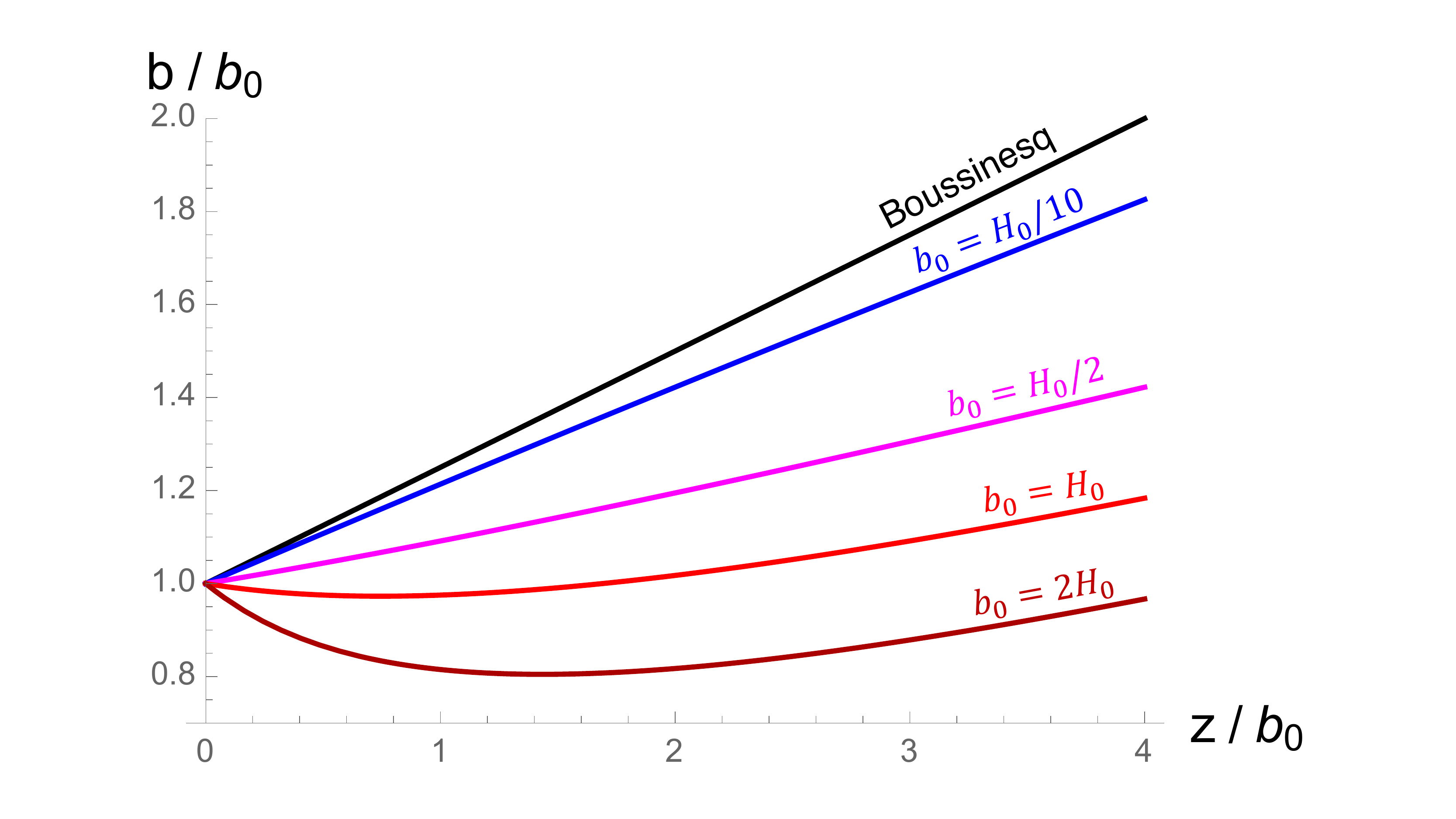}
\caption{Figure showing thermal diameter $b$ as a function of depth for different compressibility regimes.  The more compressible, the more spatially coherent this theory predicts the thermal to remain.}
\label{fig:analytic-size}
\end{figure}

We now modify these equations to account for the compressibility of an ideal gas adiabatic atmosphere. 
From hydrostatic equilibrium and the ideal gas equation of state, the density profile obeys
\begin{equation}
\rho = \rho_0 \left( 1 + \frac{\gamma-1}{\gamma} \frac{z}{H_p} \right)^{1/(\gamma-1)}
\label{eq:density-scaleheight}
\end{equation}
where $\rho_0$ is the density at $z=0$, $\gamma$ is the Gr\"{u}neisen parameter, and $H_p = k_B T_0 / \mu g$ is the corresponding pressure scale height. 
We now solve for the volume of the thermal as it propagates. 
Two processes affect the volume: entrainment increases the total mass of the thermal as it draws in fluid from its surroundings, while adiabatic compression increase its density. 
Expressed formally using $V = m/\rho$, we can write $\frac{d \ln V}{dz} = \frac{d\ln m}{dz} - \frac{d\ln \rho}{dz}$.
From Equation \ref{eq:density-scaleheight}, if we assume the pressure is equal inside and outside the thermal and that both the atmosphere and the thermal obey the ideal gas equation of state, then $\frac{d\ln \rho}{dz} = \frac{d}{dz} \left[\frac{1}{\gamma - 1} \ln \left(1 + \frac{\gamma - 1}{\gamma} \frac{z}{H_p} \right) \right] = \frac{1}{\gamma H_p + z(\gamma-1)}$. 
Therefore using Eq.~\ref{eq:entrainment}, 
\begin{equation}
\frac{d \ln V}{dz} = \frac{3 \alpha}{b} - \frac{1}{\gamma H_p + z(\gamma - 1)}.
\end{equation}
Written another way, 
\begin{equation}
\frac{db}{dz} = \alpha - \frac{b}{3 H},
\label{eq:dbdz}
\end{equation}
where $H \equiv \gamma H_p + z(\gamma - 1)$ is the $z$-dependent density scale height. 
Written this way, we can clearly see how the compressibility of the atmosphere affects the thermal's evolution. 
While entraining outside gas tends to increase the thermal mass and therefore size, the environment does work on the thermal to compress it and decrease its size. 
If $b \ll H$, then this correction can be negligible. 
However if $b \sim H$, the effect can be substantial and can even overwhelm the effect of entrainment such that $db/dz < 0$ so that the thermal shrinks rather than grows as it propagates. 
We plot the thermal size as a function of depth in Figure~\ref{fig:analytic-size}. 
We see in the Boussinesq case (black curve) that the thermal grows in size linearly as it sinks. 
When $0 < b_0 \ll H_0$, this model remains accurate, with a small correction. 
When $b_0 \sim H_0$, the thermal can propagate some distance while maintaining its size and concentration; it may even shrink and grow more concentrated when $b_0 > 3 \alpha H_0$. 
In general for the atmosphere described in Eq.~\ref{eq:density-scaleheight}, the curves shown in Figure~\ref{fig:analytic-size} are concave up because $H$ increases with depth. 
So even for initially large $b_0$, the increase in $H$ always eventually dominates and leads to a growth and dilution of the thermal. 
\\

From Eq.~\ref{eq:dbdz} we directly see why atmospheric compression might keep downwelling thermals spatially localized.  
We can inspect the problem more completely by fully solving for the equations of motion. 
We seek to modify Equations~\ref{eq:size-incompressible}--\ref{eq:buoyancy-incompressible} for a thermal falling through an adiabatic atmosphere described by Eq.~\ref{eq:density-scaleheight}. 
We write the dynamic evolution equations for $b$, $w$, and $g'$ expressed as coupled first order ordinary differential equations so that we can straightforwardly solve the system using a standard 4th order Runge-Kutta integrator. 
From Eq.~\ref{eq:dbdz}, 
\begin{equation}
\frac{db}{dt} = w \left( \alpha - \frac{b}{3H} \right).
\label{eq:size-compressible}
\end{equation}
Next we modify Eq.~\ref{eq:velocity-incompressible} to obtain an expression for $w$. 
We can use the same assumption of conservation of momentum to obtain an expression similar to Eq.~\ref{eq:velocity-incompressible}, but without the Boussinesq approximation that discards terms $\propto g'/g$ we obtain the more exact expression  
\begin{equation}
\frac{d w}{dt} = g' + \frac{\alpha w^2}{b} \left(4 \frac{g'}{g} - 3 \right).
\label{eq:velocity-compressible}
\end{equation}
Eq.~\ref{eq:velocity-compressible} is equivalent to Eq.~\ref{eq:velocity-incompressible} when $g' \ll g$. \\

Finally we modify Eq.~\ref{eq:buoyancy-incompressible} to obtain an expression for $g'$. 
One can follow the same derivation up to the final step before invoking the Boussinesq approximation, just allowing $\rho$ to vary with $p$ and assume $\frac{d\ln\rho_1}{dp} = \frac{d\ln \rho}{dp}$ as is appropriate for an adiabatic process for an ideal gas. 
Then using Eq.~\ref{eq:size-compressible}, we obtain 
\begin{equation}
\frac{dg'}{dt} = -\frac{3 w g'}{b} \left(\alpha - \frac{b}{3H} \right) \left(1 + \frac{g'}{g} \right), 
\label{eq:buoyancy-compressible}
\end{equation}
where the factor of $(\alpha - b/3H)$ arises from Eq.~\ref{eq:size-compressible} and the factor of $(1+g'/g)$ arises from relaxing the Boussinesq approximation. 
Eq.~\ref{eq:buoyancy-compressible} reduces to Eq.~\ref{eq:buoyancy-incompressible} when $b \ll H$ and $g' \ll g$. \\

Eqs.~\ref{eq:size-compressible}--\ref{eq:buoyancy-compressible} can be rewritten in their non-dimensional forms 
\begin{equation}
\frac{d\beta}{d\tau} = \omega \left(\alpha - \frac{\beta}{3} \frac{b_0}{H} \right)
\label{eq:size-dimensionless}
\end{equation}
\begin{equation}
\frac{d\omega}{d\tau} = \tilde{g}' + \frac{\alpha \omega^2}{\beta} \left(4 \tilde{g}' \frac{g'_0}{g} - 3 \right)
\label{eq:velocity-dimensionless}
\end{equation}
\begin{equation}
\frac{d\tilde{g}'}{d\tau} = -\frac{3 \omega \tilde{g}'}{\beta} \left(\alpha - \frac{\beta}{3} \frac{b_0}{H} \right) \left(1 + \tilde{g}' \frac{g_0'}{g} \right)
\label{eq:buoyancy-dimensionless}
\end{equation}
by substituting $\beta = b / b_0$, $\tilde{g}' = g'/g_0'$, $\omega = w / (b_0 g_0')^{1/2}$, and $\tau = t / (b_0 / g_0')^{1/2}$, where a 0 subscript indicates the parameter value at $\tau = 0$. 
Assuming $\omega = 0$ at $\tau = 0$, then, the subsequent dynamics are fully specified by the ratios $b_0 / H$ and $g_0' / g$. 
Of particular interest to note, when $g_0' \ll g$, then Eqs.~\ref{eq:size-dimensionless}--\ref{eq:buoyancy-dimensionless} become approximately independent of $g_0'$, because $g_0'$ only appears as a ratio with $g$. 
Thus the non-dimensional behavior of these equations of motion are approximately independent of the initial buoyancy perturbation if $g_0' \ll g$. 
As we will see in Sec.~\ref{sec:results}, this result is reproduced in our numerical simulations. \\

\begin{figure}
\centering
\includegraphics[scale=.6]{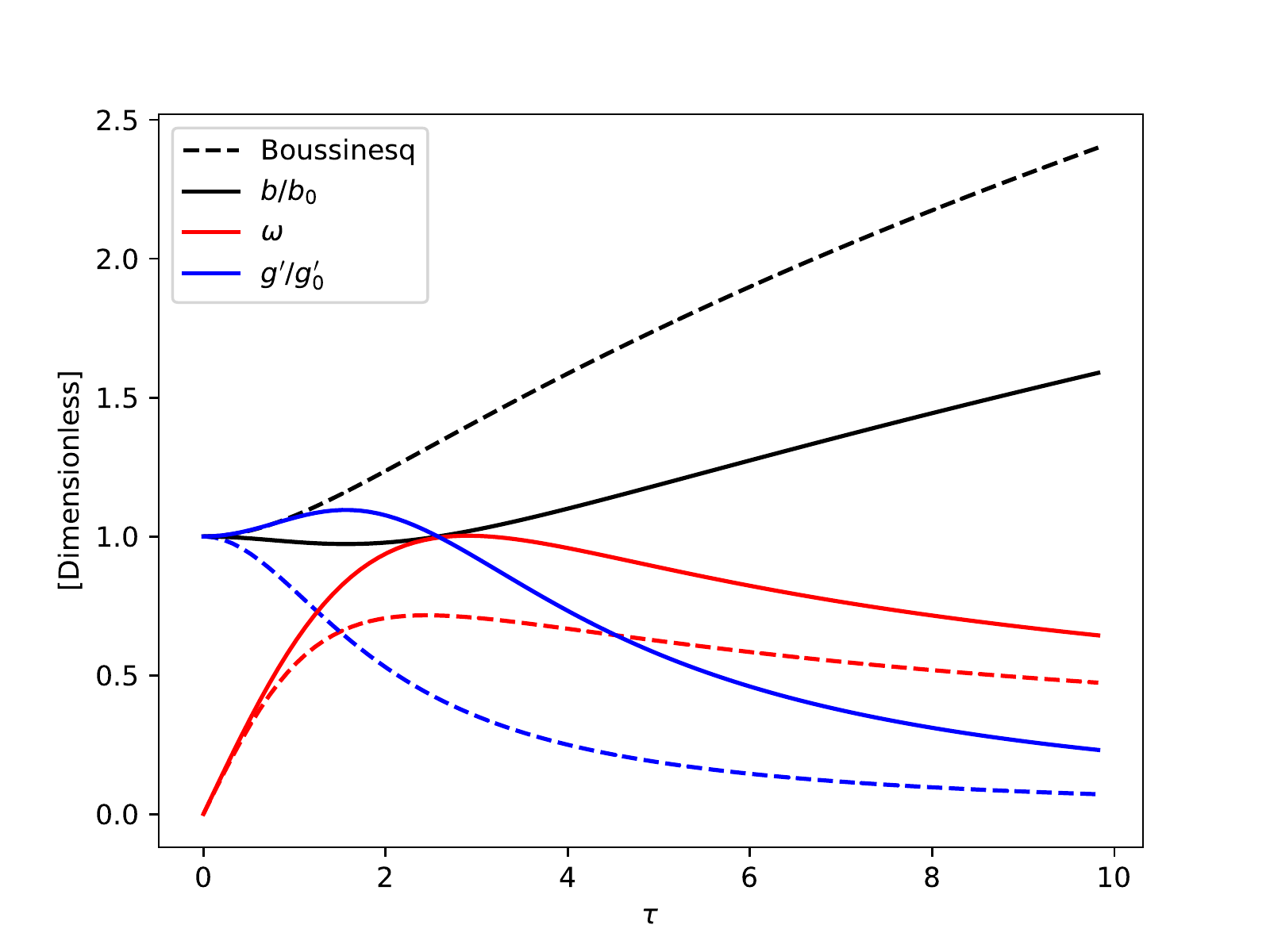}
\caption{Figure showing how the three non-dimensionalized variables of interest evolve for the uniform Boussinesq case (dashed, Equations~\ref{eq:size-incompressible}--\ref{eq:buoyancy-incompressible}) and the compressible case (solid, Equations~\ref{eq:size-compressible}--\ref{eq:buoyancy-compressible}).  
$b$ is the thermal vortex ring diameter, $\omega \equiv w/(b_0 g_0')^{1/2}$ is its non-dimensionalized mean propagation velocity, $g' \equiv \frac{\overline{\delta \rho}}{\rho_1} g$ is its mean buoyancy perturbation, and $\tau \equiv t / (b/g_0')^{1/2}$ is the non-dimensional time. 
The subscript $0$ denotes initial values. 
For this figure we use the following parameters: $b_0 / H_0 = 1$, $\omega_0 = 0$, $g'_0 / g = 10\%$, and $\alpha = 1/4$. }
\label{fig:analytic_allvariables}
\end{figure}

We show how the compressive corrections modify the behavior compared to the Boussinesq case in Figure~\ref{fig:analytic_allvariables}. 
The figure shows the evolution of non-dimensionalized parameters $b$, $w$, and $g'$ and compares the results from Equations~\ref{eq:size-incompressible}--\ref{eq:buoyancy-incompressible} against Equations~\ref{eq:size-compressible}--\ref{eq:buoyancy-compressible}. 
We see that this model predicts the thermal to remain more spatially localized, thereby enhancing the retained buoyancy density and a faster velocity. 
This finding is in agreement with \cite{anders+2019}, which also modeled downwelling thermals in highly compressible atmospheres. 
In that work they characterized ``stalling'' vs. ``falling'' regimes, distinguished by whether entrainment tended to dilute the buoyancy perturbation and slow subsidence (stalling), or whether compression dominates so that the thermal maintains its concentration and subsides at larger velocities (falling). 
According to the model we present here, the distinction between these regimes depends on the relative magnitude of the entrainment coefficient $\alpha$ and the ratio of the thermal size to density scale height $b/H$. \\

It is reasonable to estimate at which point we expect this simplified analytic model to break down. 
To assess this, we consider the relative importance of shear stress and buoyancy. 
This ratio has been canonically expressed using the non-dimensional Richardson number. 
\begin{equation}
\text{Ri} \equiv \frac{g}{\rho} \frac{\frac{\partial \rho}{\partial z}}{\left(\frac{\partial u}{\partial z} \right)^2}
\end{equation}
When Ri is large, buoyant forces are much greater than shear stresses and thus tend to organize the flow in such a way that Reynolds stresses are a small correction. 
When Ri becomes sufficiently small, shear stresses begin to dominate and the density current becomes susceptible to large-scale instability. 
In other applications, the flow will become Kelvin-Helmholtz unstable at Ri$<1/4$. 
It is therefore a reasonable first-guess to compute the Richardson number predicted by our analytic model in order to estimate the depth at which the flow may become unstable and thus break our assumption of self-similarity. \\

Using the stream function from Eq.~\ref{eq:hills-vortex}, we can estimate the relationship between the non-dimensional propagation velocity $\omega$ of the vortex ring and the corresponding shear it creates in its environment. 
The stream function is related to velocity according to 
\begin{equation}
\mathbf{v} = \frac{1}{r^2 \sin\theta}\frac{\partial \psi}{\partial \theta} \hat{\mathbf{r}} - \frac{1}{r \sin\theta} \frac{\partial \psi}{\partial r} \hat{\mathbf{\theta}}
\label{eq:v-spherical}
\end{equation}
We are interested in the shear flow between the vortex ring and its surroundings. 
In our simple model for the Hill's spherical vortex, there is no flow across the spherical boundary of diameter $b$. 
Therefore the flow at the boundary must be entirely in the $\hat{\theta}$ direction. 
Using Eq.~\ref{eq:hills-vortex}~and~\ref{eq:v-spherical}, we can write this component of the velocity outside of the vortex ring as 
\begin{equation}
v_\theta = - \left( 1 + \frac{b^3}{16 r^3} \right) w \sin \theta
\end{equation}
Then at the vortex-environment boundary, the shear flow is 
\begin{equation}
\frac{\partial v_\theta}{\partial r} = \frac{3 w}{b} \sin \theta
\end{equation}
which is maximum at the mid-plane $\theta = \pi/2$ to give us a characteristic shear of about $\frac{3 w}{b}$. 
Now we estimate the corresponding buoyancy gradient. 
If we take $\frac{\partial \rho}{\partial z} \sim \frac{\Delta \rho}{b/2}$, then $\frac{g \frac{\partial \rho}{\partial z}}{\rho} \sim \frac{2 g'}{b}$. 
This gives us an expression for the scaling coefficient of the Richardson number in terms of our analytic parameters, 
\begin{equation}
\text{Ri} \sim \frac{2 b g'}{9 w^2}
\label{eq:richardson-scaling}
\end{equation}
As argued above about Eqs.~\ref{eq:size-compressible}--\ref{eq:buoyancy-compressible}, in the limit where $g' \ll g$ the non-dimensionalized equations of motion are approximately independent of our choice of the magnitude of the initial buoyancy perturbation $g_0'$, implying that the evolution of the non-dimensional Richardson number is likewise independent of this choice. 
This suggests that the penetration depth of downdrafts is actually not sensitive to the magnitude of the buoyancy perturbation. 
As we will see in Sec.~\ref{sec:penetration}, this conceptual result is actually reproduced by our numerical simulations. \\

\begin{figure}
\centering
\includegraphics[scale=.6]{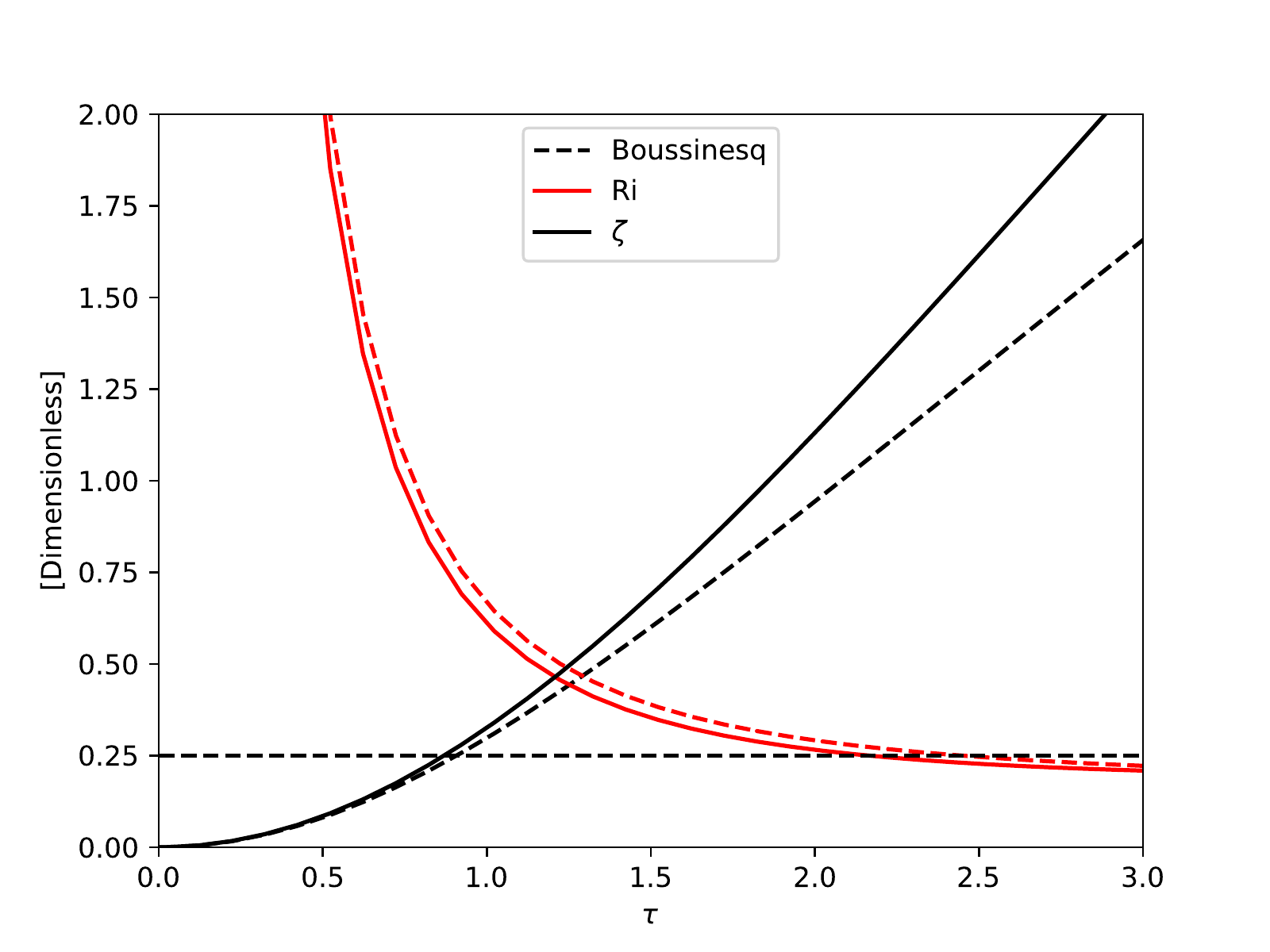}
\caption{The red curves show the Richardson number as a function of $\tau$ of the propagating vortex ring using Eq.~\ref{eq:richardson-scaling} and the computed parameters from Fig.~\ref{fig:analytic_allvariables}.  Dashed and solid curves follow the same convention from Fig.~\ref{fig:analytic_allvariables}.  The black curves show the corresponding depth $\zeta \equiv z/b_0$.  The dashed horizontal line corresponds to Ri=1/4, at which point the flow becomes susceptible to Kelvin-Helmholtz instabilities. }
\label{fig:richardson}
\end{figure}

In Fig.~\ref{fig:richardson}, we show how the behavior of the Richardson number evolves with time, and the corresponding depth of the vortex ring. 
We see that $\text{Ri} < 1/4$ around $\tau \sim 2.1$ corresponding to $z/b_0 \sim 1.25$. 
Beyond that depth, the flow may be susceptible to instability. 
Determining the nature of any such instability, as well as the timescale on which it self-amplifies to substantially disrupt the self-similar flow structure outlined here, is a complex fluid dynamical problem that requires a more sophisticated model than we have presented in this section. 
As we demonstrate in Sec.~\ref{sec:results}, the buoyant perturbation actually continues sinking to a much greater depth before finally breaking apart. 
We therefore compare the results from this simplified framework with fully hydrodynamic simulations in Sect.~\ref{sec:results}. 
We also use the analytical framework from this section as a basis to inspect additional confounding factors in Sect.~\ref{sec:confounding}. 

\section{Numerical model}
\label{sec:snap}
While simple models are useful for building intuition for complex phenomena, their efficacy must in general be compared against experimental results. 
The framework we followed in Sect.~\ref{sec:turner} was tested in a laboratory using tanks of water and colored brine \citep{morton+1956}. 
For the highly compressible environments spanning multiple density scale heights of interest to us, laboratory experiments are impractical. 
The Earth's troposphere possesses only a single density scale height when we are interested in semi-infinite atmospheres with relevant dynamics playing out over three or more density scale heights (orders of magnitude different densities).  
Even for the densest gases, building a column of multiple density scale heights would be prohibitively impractical. 
Using sulfur-hexafluoride with $\mu = 146$~g/mol, for example, building a column of three density scale heights would be $H = \gamma k_B T / \mu g \sim$~7km in height using Earth's gravity and surface temperature, taller than any manmade structure and rivaling the height of Mt. Everest. 
Furthermore it would require an immense heat source to sustain an adiabatic gradient over that distance because its adiabatic lapse rate would be far steeper than the surrounding atmosphere's. 
Given the obvious impracticality of constructing such an apparatus, we instead employ numerical models to perform computational experiments to test our model and account for the highly chaotic nature of turbulence. 
We seek to do this in a straightforward, easily reproducible way. 
However, cognizant of the possible biases that result from our assumptions and the limits of fluid dynamical simulations, we discuss confounding factors and additional tests in Sect.~\ref{sec:confounding}. \\

\subsection{Settings}
We use the flexible SNAP software \citep{snap} to simulate a localized negative density anomaly in a deep atmosphere by numerically integrating Euler's equations using a Riemann solver, with total energy as the conserved quantity. 
Because of the uncertainty in initial conditions for a storm, we assume all condensible material has been vaporized and is therefore fully miscible in its environment. 
Because it has been explicitly verified against benchmark results, and to maximize simplicity and reproducibility of our results, we opt to use the initial conditions of the 3D Straka problem \citep{straka+1993}. 
First, we initiate an isentropic, ideal gas atmosphere defined by a given temperature $T_0$ at a given pressure level and obeying Eq.~\ref{eq:density-scaleheight} with an ideal gas equation of state. 
Embedded in the atmosphere, we introduce a buoyancy anomaly centered at $z=x=y=0$ obeying 
\begin{equation}
\delta T =  \left\{
	\begin{array}{ll}
		-T_0 \frac{g'}{g} \frac{\pi^3}{6(\pi^2 - 8)} \cos \left( \frac{\pi r}{b} \right) & \quad r \leq b/2 \\
		0 & \quad r > b/2
	\end{array}
\right.
\label{eq:straka}
\end{equation}
where $r$ is the distance from the center of the bubble, and $b$ is the size of the bubble using notation consistent with Sect.~\ref{sec:turner}. 
The temperature perturbed region is isobaric at each vertical level with its environment, such that $\delta T / T = \delta \rho / \rho$. 
The choices of coefficients normalizes Eq.~\ref{eq:straka} so that definition of $\int \frac{\delta T}{T_0}dV = \frac{g'}{g}$ holds. 
Then our non-dimensionalized results are independent of our particular choices for $T_0$ and $g$ if $g' \ll g$. 
We calculate the specific entropy perturbations as 
\begin{equation}
\delta s = c_p \ln \left( \frac{T}{T_{\rm i}} \right) + R \ln \left( \frac{p}{p_{\rm i}} \right),
\label{eq:entropy}
\end{equation}
where $T_{\rm i}$ and $p_{\rm i}$ are the unperturbed initial temperature and pressure profile of the isentropic background atmosphere, and $c_p$ and $R$ are the isobaric specific heat capacity and the specific gas constant respectively. 
In Figures~\ref{fig:resolution}~and~\ref{fig:example} colors respresent entropy perturbation surface densities. 
These figures show a two dimensional representation of the three dimensional simulation by integrating the total entropy perturbation along the third dimension, i.e. $\int_Y \rho \delta s dy$ where $Y$ is the simulation domain in the $y$ dimension. 
We discuss the validity of this treatment in more detail in Sect.~\ref{sec:confounding}. \\

We seek to determine the extent to which explicit simulations agree with the idealized model from Sect.~\ref{sec:turner}. 
The compressibility of the problem can be described by the dimensionless ratio $b_0 / H_0$. 
Additionally, the timescale of the problem is set by initial buoyancy perturbation $g'$, the potential energy that generates the motion in $w_0$ as well as circulation and turbulence. 
For each simulation, we hold the atmospheric structure constant, varying only the initial conditions of the parameters. 
We therefore test a suite of choices for $b_0$ and track the subsequent evolution of the downdrafts. 
Between each simulation, we keep the relative resolution fixed so that if $b_0$ changes by a factor of two, so too does our grid spacing. 
In an incompressible environment, then, the simulation output should be scale-independent for flow of constant Reynolds number (which we discuss in more detail in Sects.~\ref{sec:viscosity} and \ref{sec:confounding}), a general property of the Navier-Stokes equation. 
We would expect in the incompressible case that the solution should be identical for constant Reynolds number if we use non-dimensional time $\tau = t (g'/b_0)^{1/2}$ and non-dimensional space $\zeta = z/b_0$. 
When comparing our results in this way, then, we can isolate the effect of compressibility $b_0/H_0$ on our simulation outputs. 
We report our results compare them to the expectations from Sect.~\ref{sec:turner} in Sect.~\ref{sec:results}. \\

\subsection{Viscosity and turbulence}
\label{sec:viscosity}
We now comment on the validity of assuming constant Reynolds number between flows of different scales. 
As we see from Eq.~\ref{eq:velocity-compressible}, the evolution of the velocity $w_0$ depends on the thermal size $b$ in a scale-dependent way. 
The true Reynolds number of our problem is enormous. 
As we will show in Sect.~\ref{sec:results}, the velocity scales can be on the order of hundreds of meters per second for length scales on the order of tens of kilometers, while the actual viscosity of gases under the relevant conditions should be tiny, $\nu \sim 10^{-5}$m$^2$s$^{-1}$. 
This characteristic Reynolds number of the problem would be of order $\rm Re = w_0 b / \nu \sim 10^{11}$. 
Fully modeling the microscopic dissipation of such vigorous turbulence in non-equilibrium flow is computationally impossible. 
Prior works \citep[e.g.,][]{anders+2019, zhou+2020} have addressed this problem by imposing an artificially low Reynolds number (or equivalently, an artificially high viscosity) to force the flow to remain laminar, or for the turbulence to remain sufficiently weak that it can be fully resolved. 
In this work, we use the standard Roe-Linearized Riemann solver to reconstruct the hydrodynamic fluxes from the volume-averaged quantities \citep{roe1981, toro2013}. 
To resolve hydrodynamic fluxes between adjascent grid cells, this method transforms the nonlinear Euler equations into quasi-linear equations using the Jacobian matrix of the flux vector fields, then solves the equations as if they were truly linear. 
This method introduces a resolution-dependent computational viscosity by effectively dissipating sub-grid scale motions. 
We therefore find resolution-dependence in our outputs that does not converge at the limits of our computational capacities using the Licallo supercomputing cluster at l'Observatoire de la C\^{o}te d'Azur. 
If we use sufficiently coarse resolution, the flow remains laminar. 
However, at finer turbulence-resolving resolutions we observe new dynamical instabilities not predicted by the analytic model (further discussion in Sect.~\ref{sec:results}). 

\subsection{$k-\epsilon$ turbulence closure model}
\label{sec:kE-methods}
We must ensure that our results are reproducible with different choices for turbulent closure. 
While imposing a global hyperviscosity stabilizes flow that should be unstable, we wanted to investigate whether a spatially-dependent eddy viscosity generated by the flow itself could produce results that are both physically realistic and fully resolvable. 
We seek a closure model of turbulence that can return resolution-indpendent results. 
Because of the nature of our problem, it would be inappropriate to use an imposed global eddy viscosity; we expect the turbulence to be time and space dependent because it is generated by the density currents themselves. 
We therefore seek a closure model that expresses these aspects of our problem. 
We chose to use a $k-\epsilon$ closure model for turbulence, following \cite{launder-spalding1972} and \cite{pope2000}. 
The $k-\epsilon$ model introduces two new dynamical variables: $k$, representing the unresolved kinetic energy of the turbulence, and $\epsilon$ representing the microscopic dissipation of turbulent energy. 
The turbulent kinetic energy is generated by shear in the flow, while the dissipation of turbulent kinetic energy is generated by the turbulence itself. 
While the details are not fully based on physics, the motivation is consistent with, for example, a Kolmogorov conception of isotropic fully developed turbulence, with turbulent flow generated at larger scales and being dissipated at smaller scales. 
The dynamical equations are:
\begin{equation}
\frac{D k}{Dt} = \frac{1}{\rho} \frac{\partial}{\partial x_j} \left[ \frac{\mu_t}{\sigma_k} \frac{\partial k}{\partial x_j} \right] + \frac{\mu_t}{\rho} \left( \frac{\partial u_i}{\partial x_j} + \frac{\partial u_j}{\partial x_i} \right) \frac{\partial u_i}{\partial x_j} - \epsilon
\label{eq:kE1}
\end{equation}
\begin{equation}
\frac{D\epsilon}{Dt} = \frac{1}{\rho} \left[\frac{\mu_t}{\sigma_\epsilon} \frac{\partial \epsilon}{\partial x_j} \right] + \frac{C_1 \mu_t}{\rho} \frac{\epsilon}{k} \left(\frac{\partial u_i}{\partial x_j} + \frac{\partial u_j}{\partial x_i} \right) \frac{\partial u_i}{\partial x_j} - C_2 \frac{\epsilon^2}{k},
\label{eq:kE2}
\end{equation}
where $\mu_t$ is the turbulent viscosity, $\sigma_k$ and $\sigma_\epsilon$ are the effective turbulent Prandtl numbers for turbulent energy and dissipation respectively, $C_1$ and $C_2$ are empirical coefficients, and subscripts on $u$ and $x$ are Einstein summation notation. 
From these quantities, $k$ which represents the mean kinetic energy of the unresolved turbulent flow and $\epsilon$ which represents its microscale dissipation, we can dimensionally reconstruct an eddy viscosity $\mu_t = k^2 / \epsilon$ which originate non-arbitrarily from the density currents of our problem. 
Importantly, we can see from Eq.~\ref{eq:kE1} that the turbulent kinetic energy is generated by inter-grid point shear---important given our interpretation from Sec.~\ref{sec:turner} that shear is the source of instability. 
This same quantity is dimensionally consistent with diffusivity, and can therefore be reasonably invoked to represent diffusion of material as well as momentum. 
We test the outcomes of two methods: a diffusive case, where the same diffusion coefficient $\mu_t$ diffuses both momentum and density differences, and the non-diffusive case, where $\mu_t$ acts as a viscosity only. 
Computationally we introduce buoyancy perturbations as temperature perturbations. 
Therefore the diffusion of buoyancy takes the form of a thermal conductivity. 
However, because of the considerable temperature gradient in the background atmosphere, we cannot conduct heat directly and must instead conduct a temperature analogue using dry static energy $E = c_p T + g z$. 

\section{Results}
\label{sec:results}

We observe the thermal to form a coherent vortex ring, initially sinking and remaing spatially concentrated. 
The subsequent behavior depends on the model assumptions.

\subsection{Low resolution validation of the analytic model}
\begin{figure*}
\centering
\includegraphics[scale=.5]{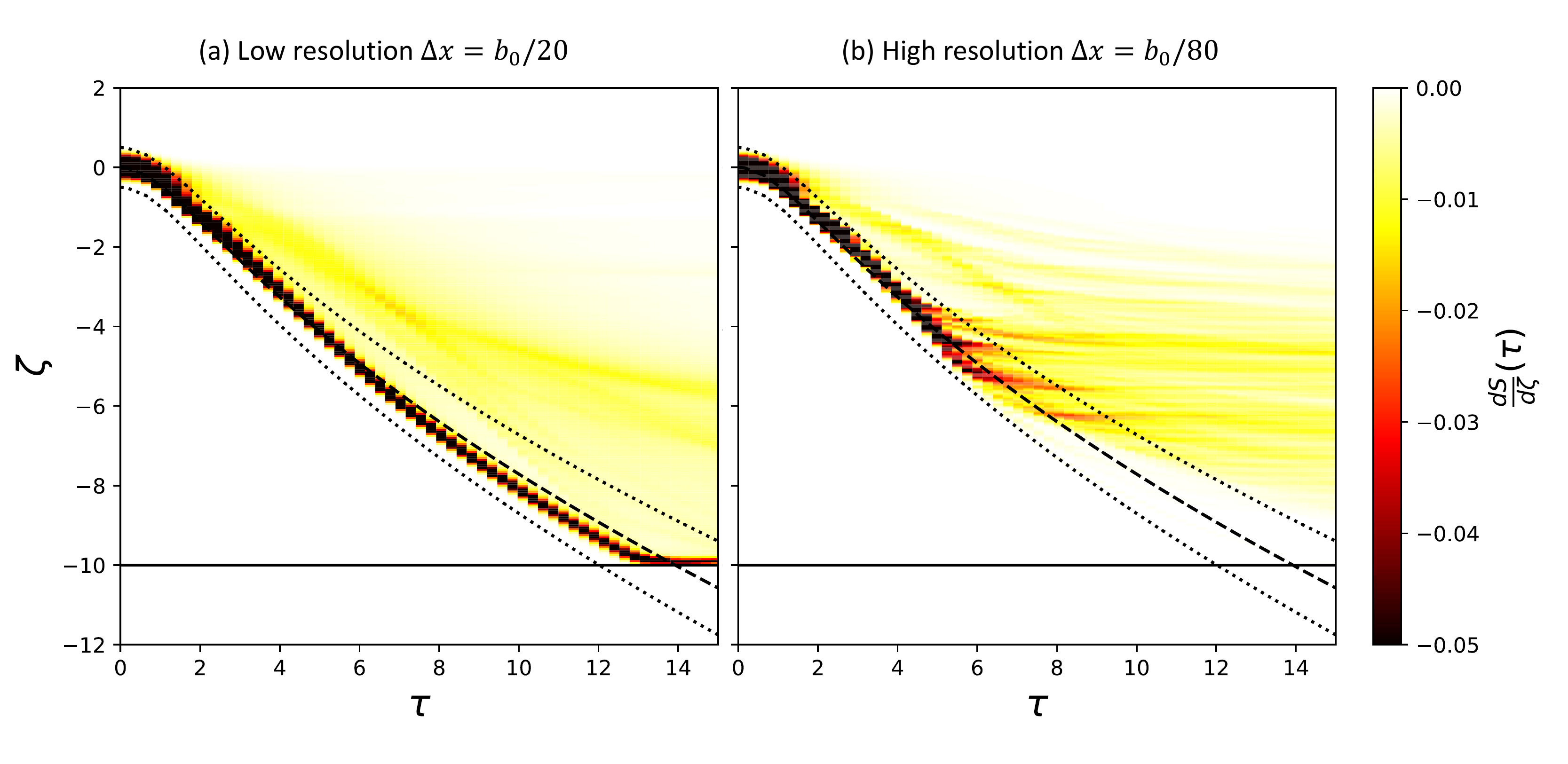}
\caption{Evolution of the initial entropy perturbation with time for different resolutions. 
$\zeta = z/b_0$ is the depth, and $\tau$ is the dimensionless time. 
Colors indicate the magnitude of the entropy perturbation at each height $\frac{dS}{d\zeta}$ according to Eq.~\ref{eq:dsdz}. 
The dashed curve shows $\zeta(\tau)$ obtained from integrating Eqs.~\ref{eq:size-compressible}--\ref{eq:buoyancy-compressible} using $b_0/H_0 = 2/5$, $\alpha = 1/4$, and $g'/g=7.5\%$, consistent with simulation parameters. 
The dashed curves show the Hill's spherical vortex boundaries $\zeta \pm b/2b_0$. }
\label{fig:evol}
\end{figure*}

At low resolutions, as argued in Sect.~\ref{sec:viscosity}, the effective Reynolds number of the problem is reduced. 
In this case, the simulation results are in fair agreement with the prediction from Sect.~\ref{sec:turner}, see Fig.~\ref{fig:evol}(a).
In this figure, we evaluate the vertical distribution of entropy at each time step according to 
\begin{equation}
\frac{dS}{dz} = \int_X \int_Y \rho \delta s dx dy
\label{eq:dsdz}
\end{equation}
where $X$ and $Y$ are the horizontal domains of the simulation in the $x$ and $y$ dimension respectively, and $\delta s$ is calculated according to Eq.~\ref{eq:entropy}. 
A vortex ring forms as expected and subsides coherently. 
Contrary to the analytic case which assumes zero detrainment, our simulation finds that some detrainment does occur. 
Nevertheless, the motion of the spatially localized central vortex ring remains accurately described by the analytic model from Sect.~\ref{sec:turner}, and the vortex ring remains concentrated down to the simulation's bottom boundary, traversing about three density scale heights. 
This result agrees with the findings of \cite{anders+2019}, which studied a similar problem with a different method. 
In the low-resolution case, our model agrees with their assumption of laminar flow, and our results are consistent.

\subsection{Dynamical instability observed at higher resolutions}
In Sec.~\ref{sec:turner}, we discussed the susceptibility of the flow to a shear instability. 
At low resolutions, no instability occurs and the flow remains well-described by the simple model from Sec.~\ref{sec:turner}.  However, 
at higher resolutions we observe a dynamical instability that arrests the subsidence of the thermal. 
We interpret this to be due to the influence of viscosity.  
At sufficiently low-resolution, small-scale motions are not well-resolved and therefore the influence of turbulence is weak (see Sec.~\ref{sec:viscosity}). 
However, at higher resolutions the flow is less artificially stabilized and we observe the dynamical instability suspected in Sec.~\ref{sec:turner}. \\

After the onset of the instability, the flow begins to rapidly detrain most of the buoyancy perturbation contained within the thermal, effectively mixing with the environment around that depth, see Fig.~\ref{fig:evol}(b). 
We refer to this depth as the ``penetration depth'' $z_p$. 
\begin{figure}
\centering
\includegraphics[scale=.4]{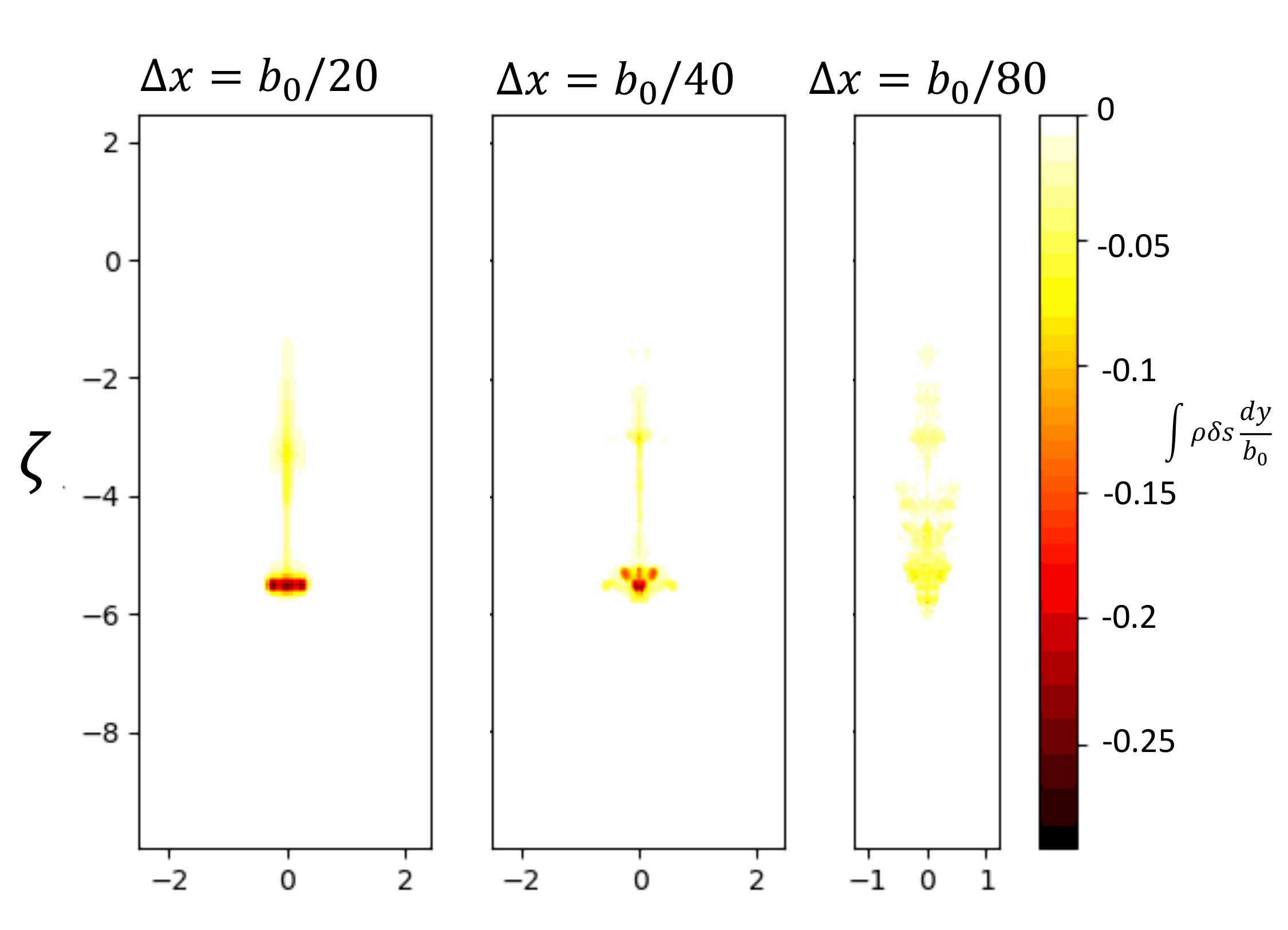}
\caption{Comparison of a snapshot in time at $\tau=8.8$ for a sinking bubble simulation with an initial size of $b_0/H_0 = 2/5$ for different choices for grid resolution, where $\zeta \equiv z/b_0$ is the dimensionless depth. 
Colors correspond to the integrated line-of-sight specific entropy perturbations, i.e., $\int_Y \rho \delta s dy$ where $\delta s$ is defined in Eq.~\ref{eq:entropy}, $y$ is the dimension coming out of the page, and $Y$ is its simulated domain. 
Units on the color bar are arbitrary, linearly proportional to entropy surface density. 
The dynamical instability observed for $\Delta x \geq b_0 / 40$ is not observed for the lowest resolution (furthest left) simulation, whose vortex ring remains stable until colliding with the bottom boundary of the simulation.}
\label{fig:resolution}
\end{figure}
In Figure~\ref{fig:resolution}, we see the higher resolution runs exhibit a dynamical instability in which much of the bubble material detrains from the primary vortex ring. 
In the low-resolution case, this instability is not observed. 
One therefore must take care to use sufficient resolution (or equivalently low enough viscosity) to capture dynamically important behaviors. 
Among models with sufficient resolution to capture the instability, the precise details of the instability are somewhat resolution dependent. 
However, for sufficiently high resolutions we observe qualitatively similar behavior, albeit with somewhat quantitatively different details. 
For our purposes in this study we are concerned with constraining the orders-of-magnitude of rainy downdrafts, and we find convergence in both qualitative behavior and order-of-magnitude penetration depth. \\

\begin{figure*}
\centering
\includegraphics[scale=.6]{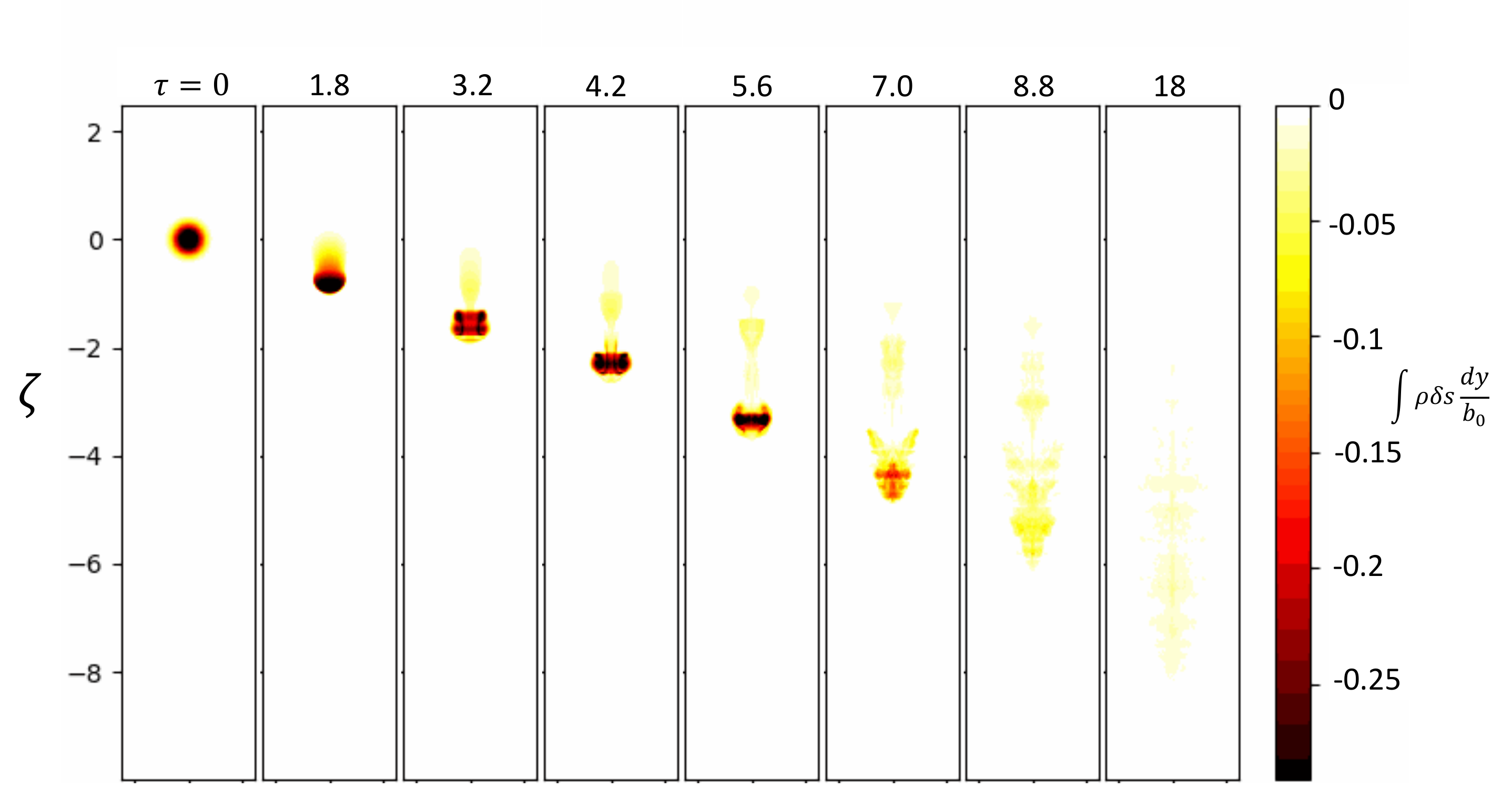}
\caption{Sample of snapshots in time for entropy perturbations of a downwelling thermal with $b_0 / H_0 = 2/5$ and $\Delta x = b_0 / 80$. 
Visualization technique is identical to Figure~\ref{fig:resolution}. 
Darker colors correspond to a larger negative entropy perturbation, integrated over the line of sight to visualize the three dimensional structure. }
\label{fig:example}
\end{figure*}

Figure~\ref{fig:example} shows the evolution of a sample simulation that initially forms a vortex ring, propagates coherently for some distance, before undergoing dynamical decoherence due to its own self-induced turbulence. 
The vortex ring instability is associated with rapid detrainment from the main thermal. 
After this, the subsidence of the centroid slows considerably. 

\subsection{Toward resolution-indpendence: the $k-\epsilon$ model}
\label{sec:kE-results}
\begin{figure}
\centering
\includegraphics[scale=.27]{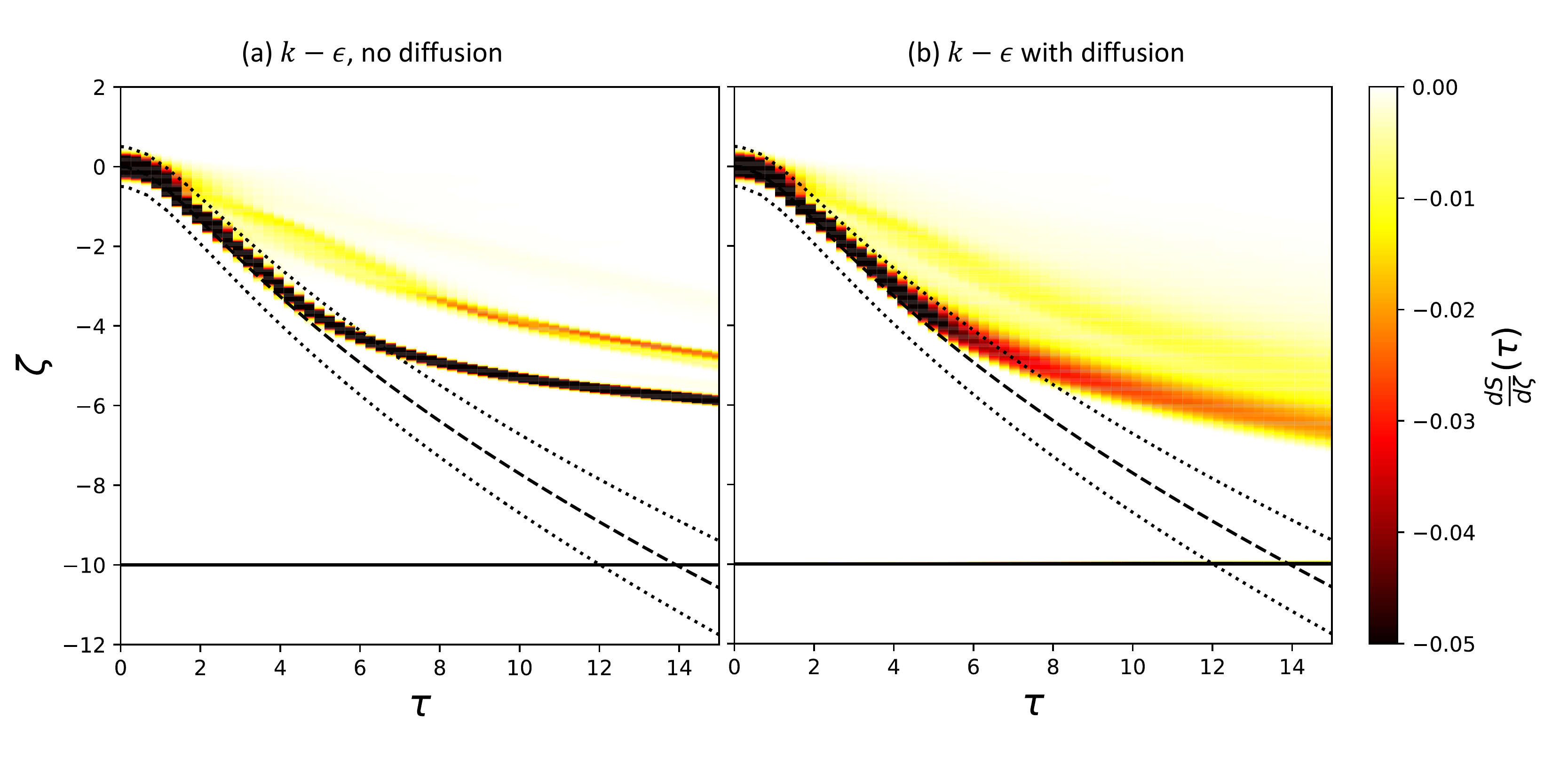}
\caption{
Sample output for the $k-\epsilon$ model, cf. Fig.~\ref{fig:evol}, using the same parameters and visualization technique. 
}
\label{fig:kE-sims}
\end{figure}

We find that the two methods for accounting for turbulence using a $k-\epsilon$ model can capture aspects of the dynamical instability for the thermal. 
The instability is characterized by both a braking in the subsidence rate dense material, and a rapid diffusion of that material. 
When we use the $k-\epsilon$ model as a pure eddy viscosity model, we reproduce the braking effect, but not the rapid diffusion of material. 
When we include diffusion of dry static energy as described in Sect.~\ref{sec:kE-methods}, the diffusion of material is better modeled, but the subsidence rate is faster than using the Roe-linearization method. 
These results can be seen in Fig.~\ref{fig:kE-sims}. 
Though imperfect, the $k-\epsilon$ method is able to reproduce these behaviors at resolutions as low as $\Delta x = b_0/10$. 
Therefore if one is interested in a particular aspect of the problem, such as finding the depth at which subsidence will appreciably slow, then the $k-\epsilon$ model can be useful. 
However, its output is not fully robust, and for more detailed calculations high resolution, turbulence resolving simulations should be used instead. 
We plot the comparison between the motion of centroid compared to the pure Riemann solver method in Figs.~\ref{fig:drops}~and~\ref{fig:compilation}; we plot the penetration depth $z_p$ (see following subsection) as a function of atmospheric compressibility $b_0 / H_0$ in Fig.~\ref{fig:penetration}.

\subsection{The penetration depth $z_p$}
\label{sec:penetration}
\begin{figure}
\centering
\includegraphics[scale=.6]{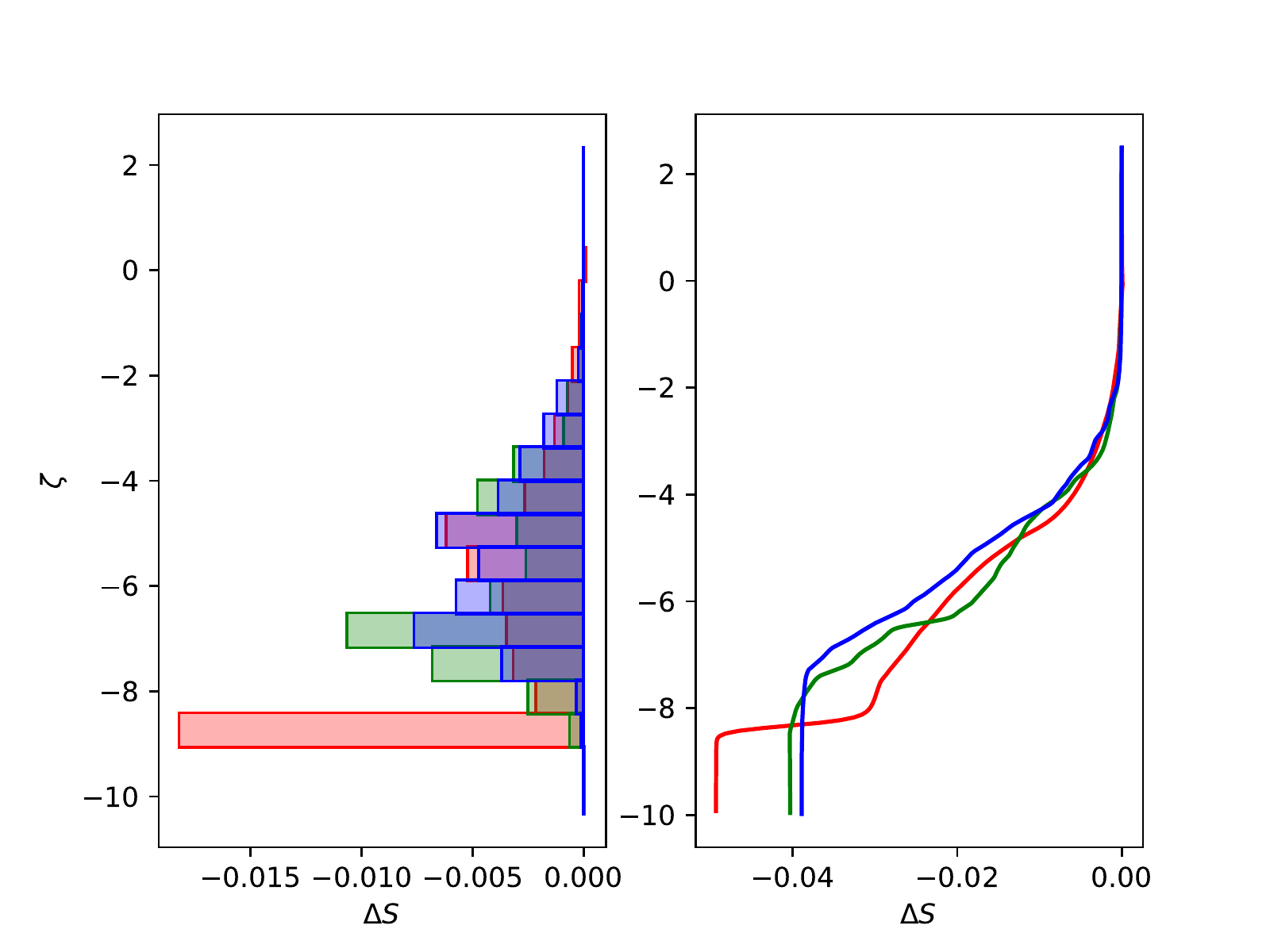}
\caption{Sample showing a histogram (left) and cumulative distribution (right, integrating from the top) of entropy at a snapshot in time for low-resolution $\Delta x = b_0/20$ (red), medium resolution $\Delta x = b_0/40$ (green), and high resolution $\Delta x = b_0 / 80$. 
$\Delta S = \int \frac{dS}{dz} d\zeta$ where $\frac{dS}{dz}$ obeys Eq.~\ref{eq:dsdz}, either integrating between bin boundaries (left panel) or cumulatively from the simulation top boundary (right panel). 
Simulation input parameters at the same as Figs.~\ref{fig:evol}--\ref{fig:kE-sims}.}
\label{fig:histogram}
\end{figure}

\begin{figure}
\centering
\includegraphics[scale=.38]{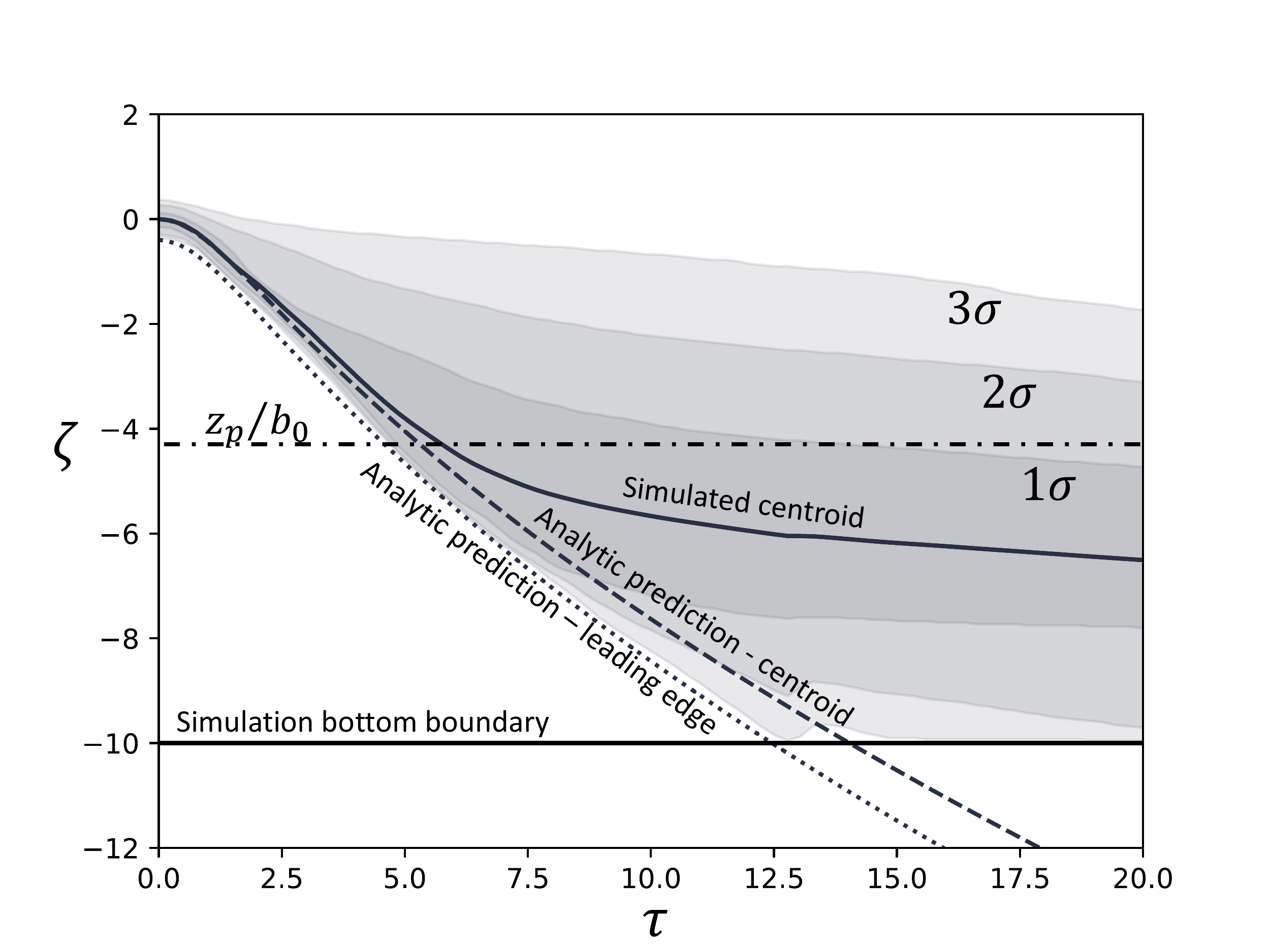}
\caption{Tracking the centroid along with the 1$\sigma$, 2$\sigma$ and 3$\sigma$ shadows representing the distribution of material for $b_0/H_0 = 2/5$ and $\Delta x = b_0 / 40$. 
The dashed curve represents the subsidence of the centroid, while the dotted line shows the analytical model prediction for the depth of the leading edge of the vortex ring from Section~\ref{sec:turner}, $z_{\rm lead} \equiv z_{\rm centroid} + b(z)/2$. 
The solid line represents the bottom boundary of the simulation. 
We use dimensionless depth $\zeta = z/b_0$ and dimensionless time $\tau = t/(b_0/g')^{1/2}$. 
The dashed-dotted line shows the penetration depth $z_p$ computed using Eq.~\ref{eq:penetration-variance} for reference.}
\label{fig:sigs}
\end{figure}

We define a penetration depth that roughly describes the depth to which most of the material comprising the initial buoyancy perturbation reaches before mixing with its surroundings. 
While there is no perfect way to describe the output of a complex hydrodynamic simulation with a single number, here we suggest a couple of quantitative methods to do so in order to quickly compare and communicate the output of simulations with different parameters. 
One method is to quantify for how long the majority of material remains spatially concentrated. 
Using Eq.~\ref{eq:dsdz}, we can bin the entropy perturbations into bins of size $\Delta z$ at any specified time. 
We can use this to plot a histogram and cumulative distribution function of entropy perturbation as shown in Fig.~\ref{fig:histogram}. 
We note that total entropy is not conserved, but rather always increases according to the second law of thermodynamics. 
Therefore $\Delta S$ will always be increasing as material mixes, and as kinetic energy is dissipated as heat. 
We note that while higher resolution runs have already largely mixed into a quasi-static distribution, the low-resolution case remains strongly peaked in a vortex ring that continues to sink rapidly (see previous sub-sections). 
Another way to visualize the simulation is to demarcate the distribution of material using contours of standard deviation, by tracking the centroid. 
Figure~\ref{fig:sigs} shows how the dense material distributes itself as the downwelling thermal propagates. 
The dashed curve shows the location of the centroid, defined such that 
\begin{equation}
\int_{z_{\rm btm}}^{z_{\rm centroid}} \frac{dS}{dz} dz = \int_{z_{\rm centroid}}^{z_{\rm top}} \frac{dS}{dz} dz
\end{equation}
where $z_{\rm btm}$ and $z_{\rm top}$ are the bottom and top boundaries of the simulation. 
The dashed curve shows the analytic prediction for the center downwelling thermal, while the solid curve shows the numerically computed simulation centroid. 
We see the analytic prediction initially agrees closely with the analytic prediction, but abruptly branches off at some characteristic penetration depth characterized by the dynamical instability visualized in Figure~\ref{fig:example} around $\tau = 7$. 
Despite this, the leading edge of the downdraft appears to remain roughly consistent with analytical predictions even after the disruption. 
The analytic prediction is shown as a dotted curve in Figure~\ref{fig:sigs}. 
Evidently, a coherent downwelling thermal continues to propagate despite having detrained a significant fraction of its initial mass. 
At late times, interaction with the bottom boundary of the simulation box becomes important. 
Figure~\ref{fig:sigs} also shows the distribution of entropy. 
If we take the distribution of entropy $\frac{dS}{dz} \Delta z$ to be a histogram with bin size $\Delta z$ according to the simulation resolution, then the shadows of different opacities show the 1, 2, and 3$\sigma$ distributions of material. 
Most of the initial entropy perturbation, then, is contained in the darkest shadow of Figure~\ref{fig:sigs}. 
We see, then, that for this particular simulation, most of the initial buoyancy perturbation penetrates to large $\zeta$; for this example most material ends up at $\zeta > 6$. 

So one quantative option could be to track the dispersion of material as the downdraft subsides, and define $z_p$ according to the point as which the dispersion of material has increased by some factor. 
While the thermal begins spatially concentrated, it spreads out as it mixes with the environment. 
For example, we can quantify the vertical distribution of material around the centroid as seen in Figure~\ref{fig:sigs}. 
We can define the penetration depth to be the point when the standard deviation of material distribution around the centroid increases by a factor of $e$. 
\begin{equation}
\sigma(z_p) = e \sigma_0
\label{eq:penetration-variance}
\end{equation}

\begin{figure*}
\centering
\includegraphics[scale=.55]{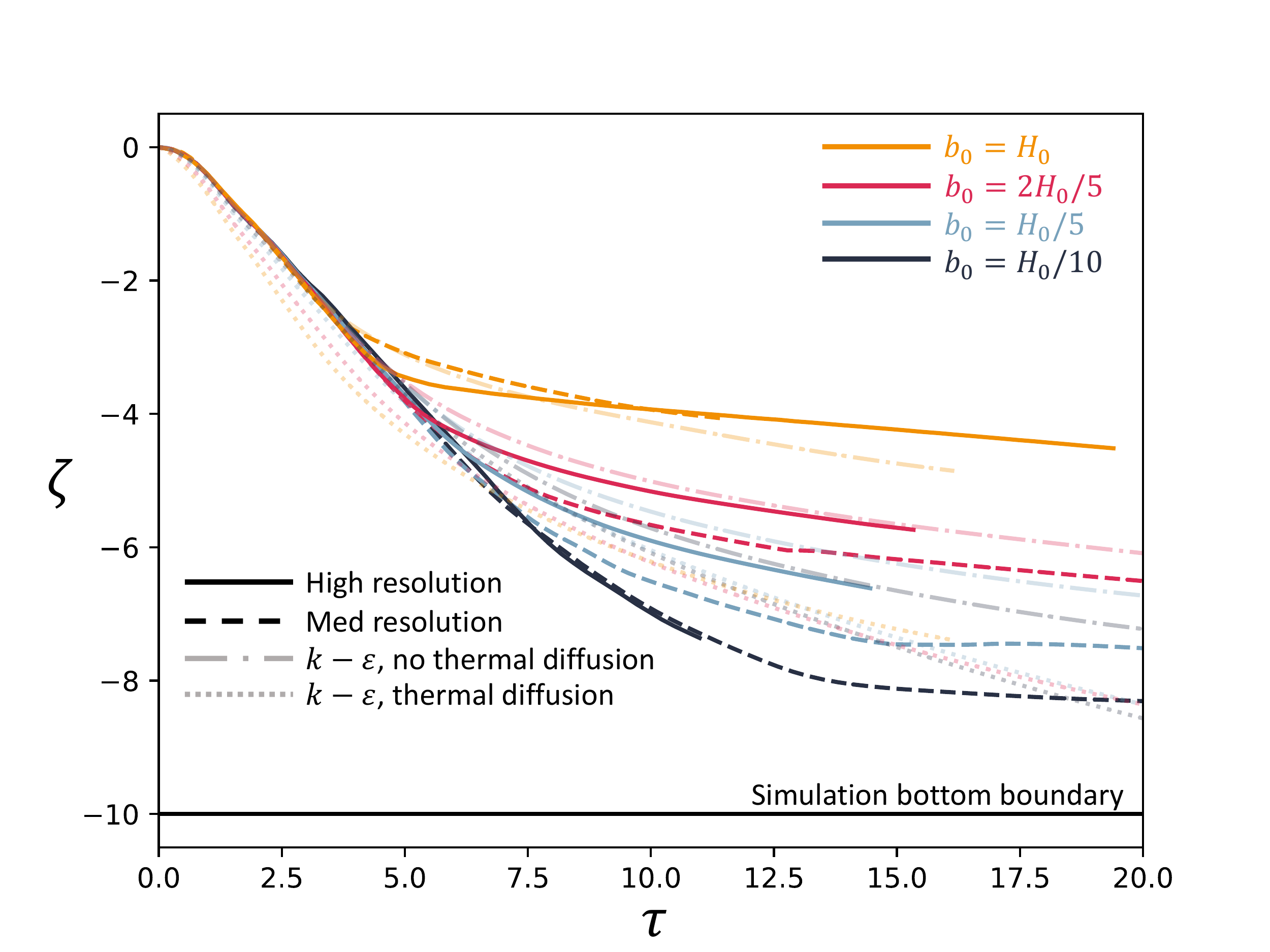}
\caption{Tracking the position of the centroid for a variety of model types, varying resolution and computational treatment of turbulence.  ``High resolution'' refers to $\Delta x = b_0 / 80$ while ``Med resolution'' refers to $\Delta x = b_0/40$.  $k-\epsilon$ models were conducted using $\Delta x = b_0 / 40$.}
\label{fig:drops}
\end{figure*}

\begin{figure*}
\centering
\includegraphics[scale=.55]{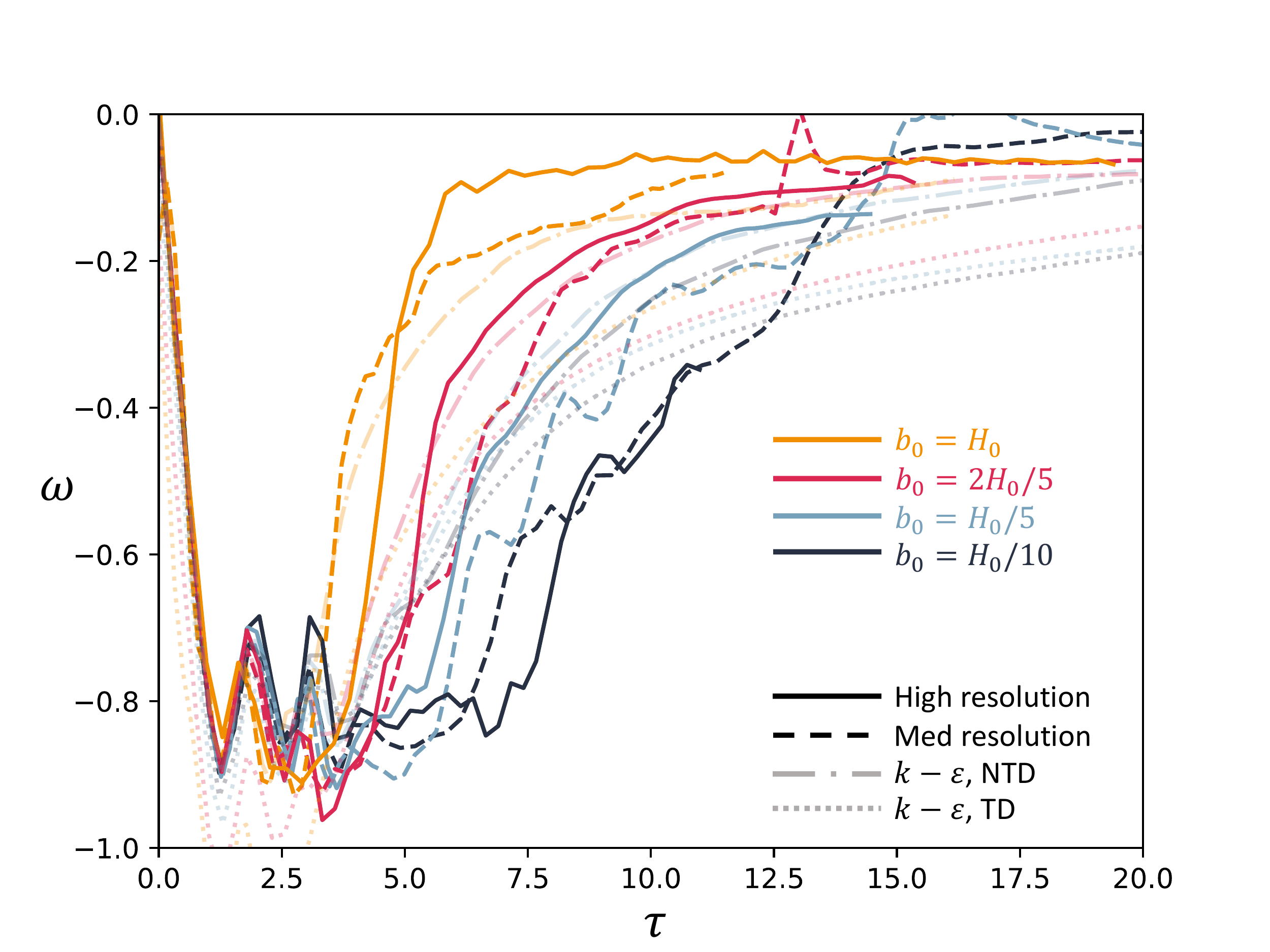}
\caption{Tracking the non-dimensional vertical velocity $\omega = w / (b_0 g'_0)^{1/2}$ of the centroid as a function of non-dimensional time across a variety of parameters using different computational methods, see Fig.~\ref{fig:drops}.}
\label{fig:compilation}
\end{figure*}

An alternative option could be to define $z_p$ to be the depth at which subsidence appreciably slows, so that the subsidence of the thermal no longer dominates compared to random turbulent motions in the surrounding environment. 
We can track the centroid of entropy perturbations for the simulation, shown in Fig.~\ref{fig:drops}. 
After the thermal vortex ring undergoes dynamical disruption, the vertical velocity of the centroid branches off of the analytic expectation, plateauing at a much lower value and mixing with the environment (see also Fig.~\ref{fig:sigs}). 
We can then compute dimensionless velocity $\omega = w/(b_0 g_0')^{1/2}$ as the temporal derivative of the cubic spline interpolation of the centroid position shown in Fig.~\ref{fig:drops}. 
The result is shown in Fig.~\ref{fig:compilation}. 
We can define the penetration depth to be the point at which the subsidence rate reduces to a value comparable to the characteristic velocity environmental turbulence, 
\begin{equation}
w_0(z_p) \sim u_0,
\label{eq:penetration-velocity}
\end{equation}
at which point we say that it can be disrupted and efficiently mixed by the environment. 
We find that both methods of evaluating $z_p$ yield comparable results when using $u_0 / (g'_0 b_0)^{1/2} \sim 3/10$. \\

\subsubsection{Dependence on initial buoyancy perturbation $g'_0$}
The behavior, both qualitatively and quantitatively, does not appear to be sensitive to the magnitude of the perturbation $g'$ for a perturbation of constant size. 
We tested $g_0'/g$ of 1\%, 2.5\%, and 7.5\%. 
We found the motion of the centroid, when non-dimensionalized, were practically identical so as to be indistinguishable when over-plotted. 
Likewise, we found the difference in penetration depths between the 1\% and 7.5\% case to be affected by <1\%.  
This may appear counter-intuitive; one might expect that a denser mass anomaly should be able to sink faster and therefore penetrate deeper than a less dense counterpart. 
However, if we assume the self-generated shear stresses to be the source of instability, then this result is unsurprising. 
A smaller buoyancy anomaly creates less rapid fluid motions, and therefore weaker shear. 
Indeed, in Sec.~\ref{sec:turner} we argued that our analytic model predicted that the results should be independent of $g'_0$ if $g'_0 \ll g$. 
We note however that our model does not consider environmental turbulence; environmental turbulence may make a difference, because it does not necessarily scale with the parameters of the problem $b_0$ and $g_0'$ (see Sect.~\ref{subsec:turbulence}). \\

\subsubsection{Dependence on compressibility $b_0 / H_0$}
We wish to inspect the behavior while varying the input parameter $b_0 / H_0$, and determine the extent to which our results depend on different choices of computational technique. 
To simplify the visualization for different choices of $b_0 / H_0$ and different computational techniques, we show in Fig.~\ref{fig:drops} the centroid tracking for these different choices, equivalent to the solid curve in Fig.~\ref{fig:sigs}. 
Then in Fig.~\ref{fig:compilation} we show the corresponding vertical velocity, essentially the time derivative of Fig.~\ref{fig:drops}. \\

The depth at which this occurs depends on the compressibility of the problem. 
These qualitative results are robust across a range of computational simulations and methods, shown in summary in Figure~\ref{fig:compilation}. 
We seek to define the penetration depth, the depth at which the velocity slows considerably and a large fraction of the thermal's buoyancy perturbation detrains from the central vortex ring. 
This process is continuous and highly complex, so there is no perfect way to quantify where this happens. 
We choose two methods to define the penetration depth as follows. 

\begin{figure*}
\centering
\includegraphics[scale=.6]{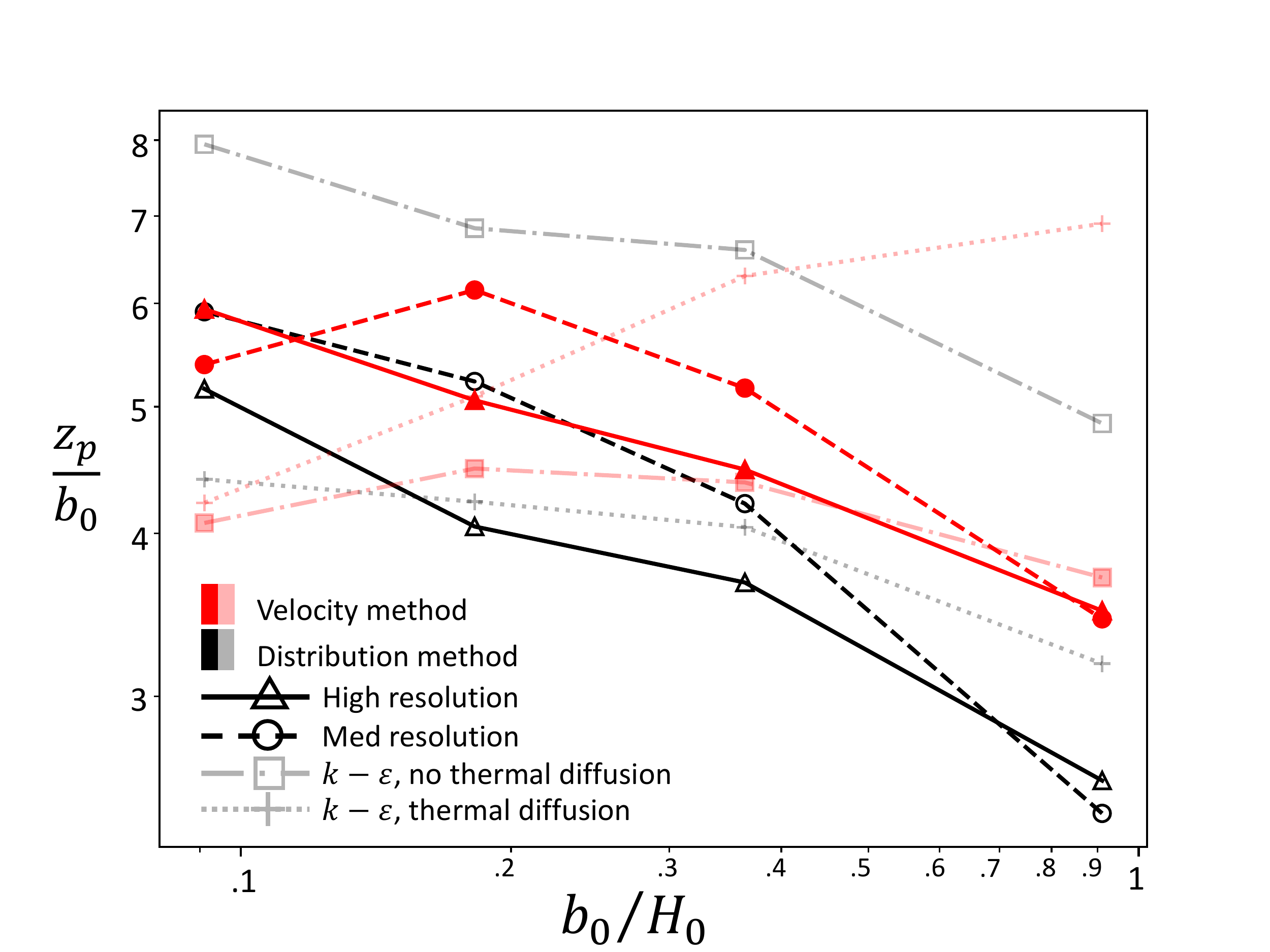}
\caption{Penetration depth of rainy downdrafts using different computational methods (see Fig.~\ref{fig:drops}), according to our definitions.  Red curves use the velocity flattening criterion Eq.~\ref{eq:penetration-velocity} with $u_0 / (g'_0 b_0)^{1/2} = 0.29$ (see also Fig.~\ref{fig:compilation}), while black curves use the vertical variance criterion from Eq.~\ref{eq:penetration-variance} (see also Fig.~\ref{fig:sigs}).}
\label{fig:penetration}
\end{figure*}

We find that our results depend somewhat on the choice for defining the penetration depth, as well as the choice of computational techniques. 
However, the topline order-of-magnitude results are consistent and generally show $z_p / b_0$ to decrease with increasing $b_0 / H_0$. 
However, the magnitude of the decrease is less than the corresponding increase in $b_0$; in particular, $z_p / b_0$ decreases only by a factor of $\sim 2$ while $b_0 / H_0$ increases by an order of magnitude (Figure~\ref{fig:penetration}). 
The results from the $k-\epsilon$ model appear instructive, if imperfect. 
The advantage of the $k-\epsilon$ model is resolution independence. 
However, the disadvantage is that only part of the dynamics are captured. 
As discussed in Sec.~\ref{sec:kE-results}, the $k-\epsilon$ model that ignores thermal diffusion, the dynamical braking around the correct $z_p$ is observed even at coarse resolutions that do not capture the instability using a standard Roe-Linearized Riemann solver. 
However, the model does not correctly reproduce the profusion of the density perturbation. 
For this reason, the $k-\epsilon$ model yields the correct $z_p$ when using Eq.~\ref{eq:penetration-velocity} but not Eq.~\ref{eq:penetration-variance} (Figure~\ref{fig:penetration}). 
On the other hand when thermal diffusion is introduced, the dynamical braking is suppressed but the profusion of material is captured; in that case, the model yields the correct $z_p$ when using Eq.~\ref{eq:penetration-variance} but not \ref{eq:penetration-velocity}. 
Therefore if one wishes to use the $k-\epsilon$ as a computational expedient for a problem like this, one must ensure it properly captures the relevant physics of interest. 
For our problem, the main takeaway is that the results are largely in agreement, regardless of these choices when accounting for the above caveats.

\section{Applications}
\label{sec:applications}
We have so far introduced an analytic and numerical framework to model rainy downdrafts in abyssal atmospheres using dimensionless parameters. 
We now seek to apply these findings to concrete celestial bodies. 

\subsection{Jupiter and Saturn}
Among our primary motivations for this study were Juno's observations of ammonia depletion below the cloud level on Jupiter \citep{li+2017}. 
We seek to estimate how our findings for dimensionless parameters from Section~\ref{sec:results} can be dimensionalized in the context of Jupiter. 
First we seek to estimate $\delta \rho / \rho$. 
In Section~\ref{sec:results} we considered these values between $\delta \rho / \rho \in [0.5\%, 7.5\%]$. 
Recall that our choice within this range does not affect the simulated penetration depth $z_p$. 
We want to estimate how the presence and vaporization of condensates affects the density of a gas parcel. 
For an ideal gas, the density is 
\begin{equation}
\rho = \frac{p \mu}{R T} = \frac{p [x \mu_c + (1-x) \mu_d]}{R T_0 - x\mu_c L}.
\end{equation}
where $T_0$ is the initial temperature of the background atmosphere before adding and vaporizing condensates, $\mu_d$ and $\mu_c$ are the mean molecular weights of the dry gas and the condensate vapor respectively, $L$ is the specific latent heat of vaporization, and $x$ is the condensate mole fraction. 
Notice that there are two ways that increasing the amount of condensate $x$ increases the density: first by changing the mean molecular weight (numerator), and second by reducing the temperature through the latent heat of vaporization (denominator). 
Then we calculate 
\begin{equation}
\frac{\delta \rho}{\rho} = \frac{1 + x\left(\frac{\mu_c}{\mu_d} - 1 \right)}{1 - \frac{x \mu_c L}{R T_0}} - 1.
\label{eq:deltarho}
\end{equation}
Now we consider two cases: ammonia condensing in the stratosphere, and water condensing a few bars beneath the photosphere around 300K. 
Considering the ammonia level corresponds to ammonia snowing as an isolated species, while the water level would imply that water and ammonia mix, perhaps as a simple fluid or perhaps following the microphysics of mushballs \citep{guillot+2020i, guillot+2020ii}. 
Starting with ammonia, we find between $3-10\times x_\odot$ (where $x_\odot$ is the solar abundance) the $\delta \rho / \rho \in (1.3\%, 4.5\%)$, comfortably within our considered range. 
For water again considering between $3-10\times x_\odot$ we find $\delta \rho / \rho \in (2.5\%, 8.5\%)$ again comparable to our considered range. \\

We now seek to constrain $b_0 / H_0$ for this problem. 
The density scale height is $H_0 = \frac{\gamma R T_0}{\mu g}$. 
For Jupiter this is about 36km at the ammonia cloud level, and 63km at the water cloud deck. 
We can estimate $b_0$ based on previous simulations that find a characteristic size of $\sim 25$km \citep{hueso+2002}. 
Fronts can be larger than individual cells, with fronts observed to be of order 50-100km wide and at times up to 400km long.  
Using 20-50km as a characteristic size range, then based on Figure~\ref{fig:penetration} we expect the penetration depth of vaporized ammonia to be roughly between 80-150km. 
Assuming the ammonia first vaporizes around the 1 bar level, this corresponds to a pressure level of $\sim 10-30$ bars. 
These findings suggest that ammonia snow on its own, if it rains from the 1 bar level as a concentrated thermal in a fashion comparable to what we describe in this paper, could plausibly penetrate to a depth of 10s of bars, as observed by Juno \citep{li+2017} without invoking any interactions with water. 
Using water we have $H_0$ that is smaller by about a factor of two, with an estimate of $b_0$ that is relatively unconstrained, because JunoCam cannot see down to the water cloud level. 
Mushball microphysics would further increase the expected penetration depth, as the vaporization of ammonia will be delayed to greater pressures. 
If this mechanism is indeed at play, it is not unreasonable to suspect that water may behave similarly. 
If we assume $b_0$ to be fixed, the penetration depth should be comparable, of order 100-150km below the water cloud deck, corresponding to a pressure level of $30-100$bars. 
Based on these arguments, it is plausible that water may likewise be depleted below its cloud deck. \\

We can use the same argument to predict that Saturn may exhibit a similar depletion in water and ammonia below their respective cloud decks. 
However, the situation on Saturn may be more complicated; Saturn's quasi-periodic Great White Spot erupts roughly every thirty Earth-years rather than quasi-continuously as on Jupiter.  
If a significant fraction of Saturn's moist convective activity occurs intermittently on multi-decadal timescales, that leaves long quiescent times during which water vapor can diffuse upward again. 
Convective inhibition may also be at play on Saturn \citep{li-ingersoll2015}, further complicating the story. 
\\

\subsection{Uranus and Neptune}
Because Uranus is a high priority target for a future Solar System mission, we consider the implications of our findings there, and on its Solar System cousin Neptune. 
We consider methane, for which we have some constraints on abundance, estimated to be of order $\sim 40\times x_\odot$ on both Uranus \citep{lindal+1987} and Neptune \citep{conrath+1991}. 
Using these numbers and taking the methane cloud layer to be at 100K, we obtain an estimate for $\delta \rho / \rho$ at least 40\%, much larger than on Jupiter. 
Therefore we can expect that perhaps downdrafts associated with such extreme methane enrichment maybe denser and stronger than the downdrafts on Jupiter, and additional considerations in violation of our assumption that $g' \ll g$ for Equations~\ref{eq:velocity-compressible}--\ref{eq:buoyancy-compressible}. 
For the following calculations will will extrapolate our finding that the penetration depth is insensitive to $\delta \rho / \rho$. 
Although methane storm activity has been observed on Uranus \citep{depater+2015} and Neptune \citep{molter+2017}, there have not been close observations of such events with the resolution and detail of JunoCam. 
If we assume $b_0 / H_0 \sim 1$, comparable to Jupiter, and calculating the density scale height on Uranus at the methane cloud level to be around 50km, we calculate the penetration depth to be 150-200km for methane storms, corresponding to a pressure level of 20-40 bars. 
The numbers would be the same on Neptune, owing to Uranus and Neptune's nearly identical effective temperatures and similar surface gravity. 
The possibility of a convectively inhibited atmosphere on Uranus and Neptune with a statically stable radiative region \citep{markham-stevenson2021}, however, may complicate the dynamics compared to the neutrally stable case investigated in this work. 
Methane has been unambiguously detected in the atmosphere of Uranus and Neptune, but if methane is depleted below the cloud level like ammonia is on Jupiter, then the abundances that have been measured may not be representative of its bulk abundance. 
Additionally, methane may not be well-mixed and could include compositional gradients \citep[cf.][]{sromovsky+2011}. 
Indeed, there is already evidence for latitudinal variation in the distribution of methane on Uranus \citep{molter+2021} and Neptune \citep{tollefson+2021}. 
Whether methane is substantially depleted to tens of bars beneath its cloud level may be tested as part of an upcoming mission. 
If its structure resembles Jupiter, that observation would favor a fluid dynamical explanation for the observed depletion.

\subsection{The Sun}
\label{sec:applications_sun}
Convection on the Sun is somewhat simplified by the absence of condensates undergoing phase transitions. 
We follow \cite{anders+2019} to estimate the relevant buoyancy and length-scales of thermals from the solar photosphere. 
For a background atmosphere of temperature $T_0 \sim 6000$K, the downflow temperature deviation is expected to be $\delta T \sim -500$K so that $\delta \rho / \rho \sim 8.3\%$, roughly consistent with our simulations. 
The thermal perturbation size should be close to the width of the thermal downflow lanes $\sim 100$km. 
For solar gravity at 274~m~s$^{-2}$ for monatomic hydrogen and helium, this implies $b_0 / H_0 \sim 1$. 
Therefore we expect the penetration depth of ``entropy rain'' to be 300-400km, or $< 0.2\%$ of the total depth of the convective zone. 
Therefore in contrast to the findings of \cite{anders+2019}, we find that atmospheric compression alone is not a sufficient mechanism to preserve the coherence of a downwelling thermal. 
The difference between our results stems from our different treatment of turbulence. 
\cite{anders+2019} considered the laminar case; in that case, our results agree. 
However when resolving turbulence using the Roe-Linearized Riemann solver method at high resolution, or using a turbulence closure model like $k-\epsilon$, the density current is arrested at a finite penetration depth $z_p$ that prevents the downwelling thermal from maintaining its coherence through the convective envelope of the Sun. 
In order for convective heat transport by entropy rain to persist \citep[as described by][]{brandenburg2016}, some other focusing mechanism, \citep[perhaps magnetic fields, see e.g.,][]{cattaneo+2003, weiss+2004}, or different assumptions about the nature of the problem would be required. 

\subsection{Exoplanets}
Beyond our own Solar System, the basic fluid dynamics we describe here should also apply to exoplanets. 
The specific length-scales of relevance would depend on the details of the system, and we will not attempt to make exhaustive quantitative predictions here. 
However, our non-dimensionalized results from Section~\ref{sec:results} can be straightforwardly applied to an exoplanetary system using the methods from this section. 
Furthermore, we note that our mechanism for depleting volatile vapor below the cloud level should serve as a caution not to mistake measured atmospheric abundances of volatile species for bulk interior abundances in planets with gas envelopes.

\section{Confounding factors}
\label{sec:confounding}

Our main analysis has employed a highly simplified picture of rainy downdrafts, aimed at identifying some of the main qualitative features of this type of flow and constraining the order-of-magnitude of downdraft penetration depth. 
The simplified assumptions employed were used as a starting point to minimize the number of free parameters that could affect our results, as well as to maximize the simplicity of reproducing our results. 
However, we must inspect whether our idealized results can be reasonably applied to real planetary environments. 
In this section we address a variety of concerns one might pose. 

\subsection{Initial conditions}
\begin{figure}
\centering
\includegraphics[scale=.4]{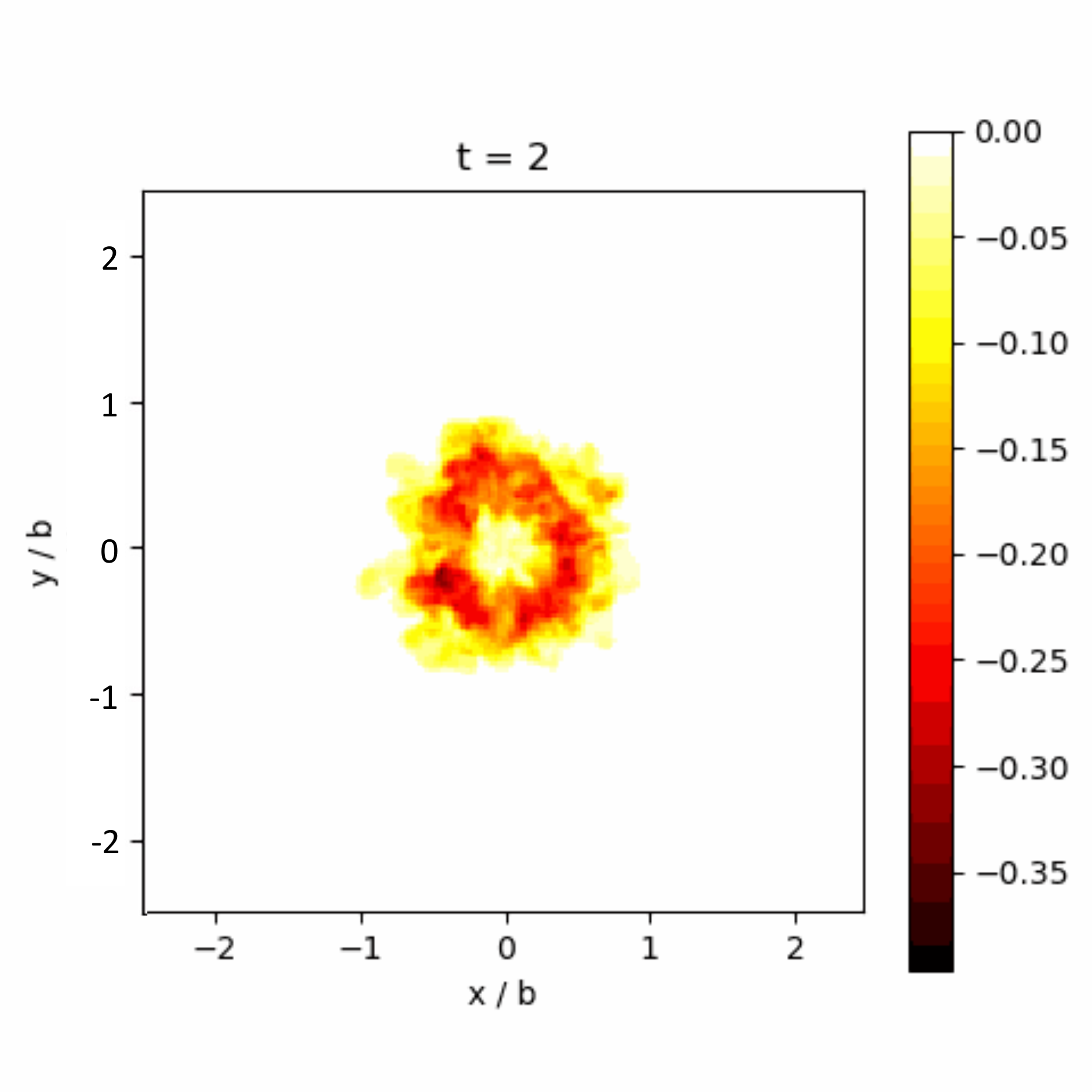}
\caption{Integrated line of sight entropy perturbations after some initial evolution of the random cube model with a top-down view, using the same plotting technique as e.g., Fig.~\ref{fig:example}. 
We observe that despite a lack of spherical symmetry upon initialization, that a vortex ring as approximately described in Sec.~\ref{sec:turner} spontaneously forms, albeit retaining some of the stochasticity from the initial conditions. }
\label{fig:cube-ring}
\end{figure}

In our computations, we chose somewhat contrived initial conditions based on the 3D Straka problem. 
One might reasonably ask whether our results would differ if some alternative initial conditions were specified. 
To inspect this problem, we compare our results for the 3D Straka case to a different arbitrary initialization. 
As one example, we begin with identical initial conditions to the 3D Straka problem, but use a uniform $\Delta T$ in lieu of Eq.~\ref{eq:straka}. 
As another example, we initialize a cube of material with density perturbations distributed randomly in a cube instead of a centrally condensed bubble as in the Straka case. 
The cube model in particular is notably different in that its initialization does not involve spherical symmetry. 
For each grid space, we assign the temperature perturbation to lie randomly between 0K and 30K over a cube sized such that the expectation value for the total entropy perturbation is comparable to the 15K Straka bubble. 
We find that despite the initialization of the random cube model bearing no spherical symmetry, a vortex ring will still spontaneously form (see Fig.~\ref{fig:cube-ring}). 
Moreover, we find that the top-line qualitative and order-of-magnitude results hold---the new initial conditions likewise allow the initial perturbation to sink down many multiples of its initial size before substantially mixing with its environment. 
The penetration depth increases by less than 30\%, which is of small consequence in the context of our order-of-magnitude interests. 
We show how penetration as measured by Eq.~\ref{eq:penetration-variance} changes with different initializations and methods in Table~\ref{tab:penetrations}. 
\begin{table}
\caption{Calculated penetration depth for different model types and initializations.  }
\begin{tabular}{c|ccc|cc}
\toprule
 & \multicolumn{3}{c}{Straka} & \multicolumn{2}{c}{Alternatives} \\
 \cmidrule(lr){2-4} \cmidrule(lr){5-6}
                   & Med. & High & $k-\epsilon$ & Uniform  & Random  \\
                   & res. & res. &  & sphere & cube \\
             \midrule
$z_p / b_0$        & 4.215   & 3.668  & 4.044        & 5.272   & 5.420       \\
$\delta z_p / z_p$ & ---       & -13\%  & -4\%         & +25\%   & +29\%  \\
\bottomrule
    
\end{tabular}
\label{tab:penetrations}
\end{table}
We compare the different numerical methods for the Straka problem medium resolution, high resolution, and the $k-\epsilon$ model also plotted on Fig.~\ref{fig:penetration} to results from simulations with different initial conditions.
For different initial conditions we investigate the uniform sphere and random cube case described above. 
While we don't find perfect agreement, we do find consistent orders-of-magnitude, indicating that our overall findings are not catastrophically sensitive to initial conditions. 
Of course, the choice of initial conditions will affect the subsequent dynamics, but our top-line results are not too sensitive to different arbitrary choices. 
Modeling a realistic storm, including the development of upwelling convective columns, the microphysics of condensation and coagulation, the precipitation and vaporization, in highly compressible environments is the subject of ongoing and future work. 
For the moment, our results apply to spatially localized downwelling thermals. 

\subsection{Environmental turbulence}
\label{subsec:turbulence}

We have considered a neutrally stratified adiabatic background atmosphere, but real planetary atmospheric dynamics is driven by cooling from the top. 
The ensuing turbulent convection that we expect, as well as driven thermal winds, shear between belts and zones, and other dynamic processes in the atmosphere should excite significant turbulence in the environment external to the downwelling thermal. 
We negleted this effect in our prior analysis. 
We do not attempt to model such an uncertain and complex phenomenon in our simulations, but we do point to theoretical and experimental work that has been done in the past. \\

\subsubsection{Disruption of thermal by turbulent surroundings}
\begin{figure}
\centering
\includegraphics[scale=.6]{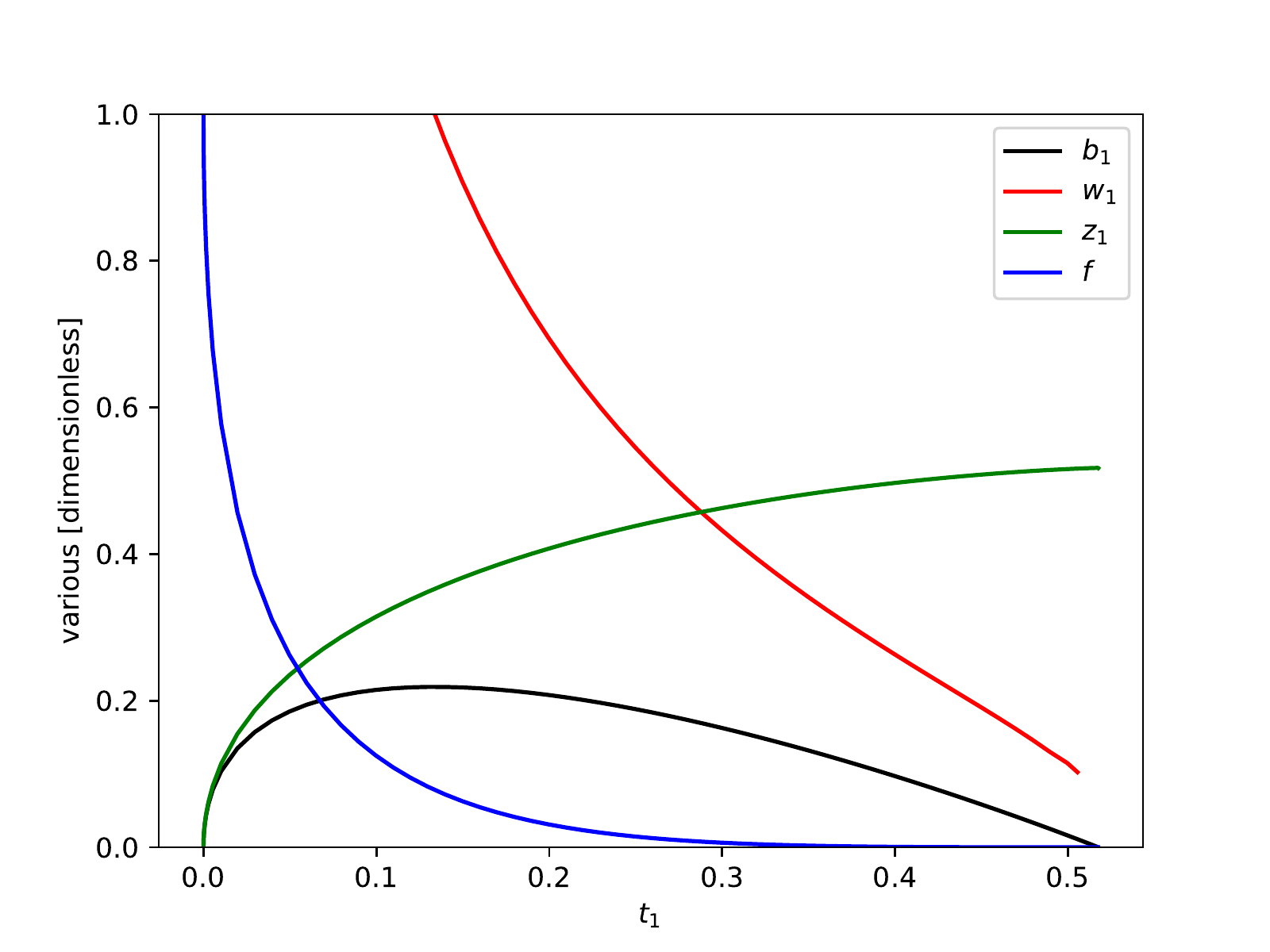}
\caption{Evolution of dimensionless parameters for a thermal propagating in a turbulent environment \citep[cf.][]{turner1962}.
$b_1 \equiv b u_0/F_*^{1/2}$, $z_1 \equiv z u_0/F_*^{1/2}$, $t_1 \equiv t u_0^2/F_*^{1/2}$, $w_1 = w/u_0$, and $f= b^3 g' / F_*$ with $F_* \equiv b^3 g_0'$. 
The simulation truncates when $b_1 \rightarrow 0$, indicating that its entire initial buoyancy perturbation has detrained into the environment. 
}
\label{fig:turb}
\end{figure}

Following \cite{turner1962}, we can model environmental turbulence in the following way. 
In the quiescent environment, the flow is driven by the motion of the thermal itself, so that entrainment is proportional to its propagation speed $w$. 
In a turbulent environment, there exists flow driven by the motion of the thermal, but also environmental flow driven by some outside dynamics. 
If the turbulence is homogenous and has characteristic length scales that are small compared to the thermal, we can model its effect on the thermal to be detrainment of material from the thermal into the environment. 
We can specify a characteristic turbulent velocity $u_0$. 
In this case, Equations~\ref{eq:size-incompressible}-\ref{eq:buoyancy-incompressible} become 
\begin{equation}
\frac{db^3}{dt} = 3\alpha b^2 w_0 - 3 b^2 u_0
\end{equation}
\begin{equation}
\frac{d (b^3 w_0)}{dt} = \frac{2}{3} b^3 g' - b^2 u_0 w_0
\end{equation}
\begin{equation}
\frac{d (b^3 g')}{dt} = - b^2 u_0 g'
\end{equation}
The above equations modify the evolution equation for the thermal diameter $b$ to be $b-b_0 = \alpha z - u_0 t$, where $z$ is the depth. 
Introducing the scaling parameter of buoyancy flux $F_* \equiv b^3 g'_0$, we can non-dimensionalize the above equations so that 
\begin{equation}
b_1 = \alpha z_1 - u_0 t_1
\end{equation}
where $b_1 \equiv b u_0/F_*^{1/2}$, $z_1 \equiv z u_0/F_*^{1/2}$, and $t_1 \equiv t u_0^2/F_*^{1/2}$. 
We can non-dimensionalize the other quantities using these same scaling parameters: $w_1 = w/u_0$, and $f= b^3 g' / F_*$. 
The evolution of this system is shown in Figure~\ref{fig:turb}.\\

Initializing $b_0 = 0$ for a point source, the thermal initially grows, before reaching a maximum size and beginning to shrink. 
At this point, the environmental detrainment begins to overwhelm the thermal entrainment rate, i.e. $w \sim u_0$. 
The thermal will continue to propagate as it shrinks, before fully disappearing at some height. 
We can solve for this numerically, finding $z_1 \sim 0.52$. 
Notice that we use $b_1(t_1) = 0$ here in contrast to $b(\tau=0) = b_0$ used in the prior sections. 
This difference arises because the non-dimensionalization in this problem uses $u_0$ and $F_*$, in contrast to our use of $g_0'$ and $b_0$ in prior sections. 
This difference arises due to the nature of the problem, because $u_0$ is a new fundamental scaling relationship for the problem. 
The statement that $b_1 \rightarrow 0$ is essentially saying that $(b_0 u_0)^2 \ll F_*$ in the beginning, i.e., the gravitational potential energy of the thermal is much greater than the turbulent kinetic energy of the surroundings.  \\

We now must estimate this quantity for a planet like Jupiter. 
For a thermal of size $b=40$km, then we expect the penetration depth when accounting for environmental turbulence to be 
\begin{equation}z_p \sim 4\times 10^3 \left( \frac{u_0}{1\text{m~s}^{-1}} \right)^{-1} \left( \frac{\overline{\Delta T}}{15 \text{K}} \right) \rm km.
\end{equation} 
Using $u_0 \sim 1$m~s$^{-1}$ is consistent with what we expect for equilibrium turbulent convection. 
However, one can use a larger value to account for meterological activity to obtain a shallower penetration depth. 
Furthermore, if $\overline{\Delta T}$ is smaller this likewise would decrease the penetration depth. 
When we compare this value and scaling to our findings using simulations scaled to Jupiter, we find the depth dictated by expected environment turbulence to be one or two orders of magnitude larger that the depth dictated by self-induced turbulence. 
Evidently, for weak environmental turbulence the effect of instability wrought by the flow itself is more important than the effect of environmental turbulence. 
Still, for highly turbulent environments or small initial buoyancy perturbations, the effect of environmental turbulence may become important. 
We therefore add this possibility as a caveat to our findings from Sect.~\ref{sec:results}. \\

We have not modified the behavior to account for the compressibility of the environment. 
We will, however, comment on the behavior we expect qualitatively. 
Following Figure~\ref{fig:analytic_allvariables}, we see that in a compressible environment we expect faster propagation velocities. 
Since the dynamics of the thermal in a turbulent environment are governed by the relative scale of $w$ and $u_0$, in general for downwelling thermals in compressive environments we expect $w$ to dominate to a deeper depth in the analytic case. 
We do not endeavor to model compressive convection in a turbulent environment fully in this work, as we have just demonstrated that environmental turbulence should be less important than self-induced turbulence for an environment like Jupiter's. 
Atmospheric compressibility does not change this conclusion, and we therefore continue to emphasize our main result that finds a penetration depth caused by a self-induced turbulent instability rather than primarily by mean environmental turbulence.

\subsubsection{Redistribution of material by eddy diffusion}
In this work, we primarily focused on one rather straightforward problem: the dynamics of an initially localized thermal propagating through a neutrally stratified atmosphere at rest. 
However in reality, such events would only be one part of the much more extensive dynamics at play in the atmosphere. 
A fuller picture of the atmosphere would include our results as one component of a larger atmospheric model. 
As one example, using the eddy-diffusion mass-flux (EDMF) models that have been applied to the Earth \citep[e.g.,][]{siebesma+2007, suselj+2019}, we have investigated plumes in one direction, but have not really discussed eddy-diffusion. 
While a fully robust atmospheric model of this kind is beyond the scope of this work, we can make some comments on the applicability of this work to future endeavors to incorporate our results into more general models of atmospheric circulation. 
EDMF models employ a modified diffusion equation of the form 
\begin{equation}
\frac{\partial \bar{\phi}}{\partial t} = \frac{-\partial}{\partial z} \left[-K \frac{\partial \bar{\phi}}{\partial z} + M (\phi_{\rm NL} - \bar{\phi}) \right]
\label{eq:edmf}
\end{equation}
where $\phi$ refers to the quantity of interest (for example the concentration of a volatile species), $\bar{\phi}$ refers to the mean abundance at a given vertical coordinate, and $\phi_{\rm NL}$ refers to non-local mixing from strong updrafts, or downdrafts in our case. 
Solving this equation in general requires some model for the mass flux $M$, as well as a relationship between $\phi_{\rm NL}$ and the propagation velocity. 
Our results from Sec.~\ref{sec:turner} as well as Fig.~\ref{fig:compilation} provide some constraints that can be used for these models. 
In addition, a current complication when endeavoring to employ EDMF models on giant planets is the lack of an obvious bottom boundary. 
While the top boundary of the stratosphere holds for giant planets as well, 
on models of the Earth, the Earth's surface acts as an obvious bottom boundary. 
On Jupiter, no such bottom boundary exists. 
Therefore when modeling plumes in abyssal environments, one must have some estimate for the distance over which it can be treated according to the analysis in Sec.~\ref{sec:turner}, and at which point the flow may become unstable and simply mix with its environment. 
Fig.~\ref{fig:penetration} offers reasonable estimates to use when constructing such a model. 
\\

We now seek to estimate how efficiently rainy downdrafts can remove volatile species, compared to the efficiency of large-scale turbulent mixing. 
If we think of the atmosphere under an eddy-diffusion paradigm, then retaining a steady-state compositional gradient requires sources and sinks of composition to accomodate the eddy diffusion flux. 
According to Fick's Law, the steady-state flux is $J \sim -K \frac{\partial \phi}{\partial z}$, using the same definition of $\phi$ from Eq.~\ref{eq:edmf}. 
If we regard the composition sink to be rainout at the top boundary, and the composition source to be the penetration of rainy downdrafts at the bottom boundary, then we can estimate the efficiency of this exchange using a simple diffusion model. \\

Consider Jupiter. 
Juno observed a compositional gradient to vary between about 100ppm at 1 bar to 373ppm at 50-60 bars \citep{li+2017}, corresponding to a distance of approximately 60km. 
The mean compositional gradient would thus be $\sim 4.5$ppm/km. 
If we use a characteristic eddy-diffusivity for Jupiter of $2 \times 10^4$m$^2$s$^{-1}$ by using Jupiter's scale height of 20km and vertical convective velocities of order 1m/s \citep{guillot+2004}, then we estimate this eddy diffusion flux to be of order $3\times 10^{-7}$ moles of NH$_3$ per square meter per second around the 1 bar level. 
Integrated over the surface of Jupiter, this corresponds to around $10^8$kg of ammonia per second. 
One plume as explored in this model, for example the nominal model of $\Delta T \sim 15$K with $b=40$km corresponds to about $2.4 \times 10^9$kg of ammonia around the 1 bar level. 
Therefore to an order-of-magnitude based on these rough numbers, we would expect to need such a plume to descend about once every four minutes somewhere on Jupiter in order to accommodate the observed compositional gradient. 
This estimate is consistent with Juno's observed lightning statistics, detecting large lightning strikes on approximately the same intermittency timescale \citep{brown+2018}. \\

This work can be regarded as a modest first step toward building a more sophisticated atmospheric model, for example and EDMF model, for Jupiter's atmosphere.

\subsection{Vertical wind shear}
\begin{figure}
\centering
\includegraphics[scale=.38]{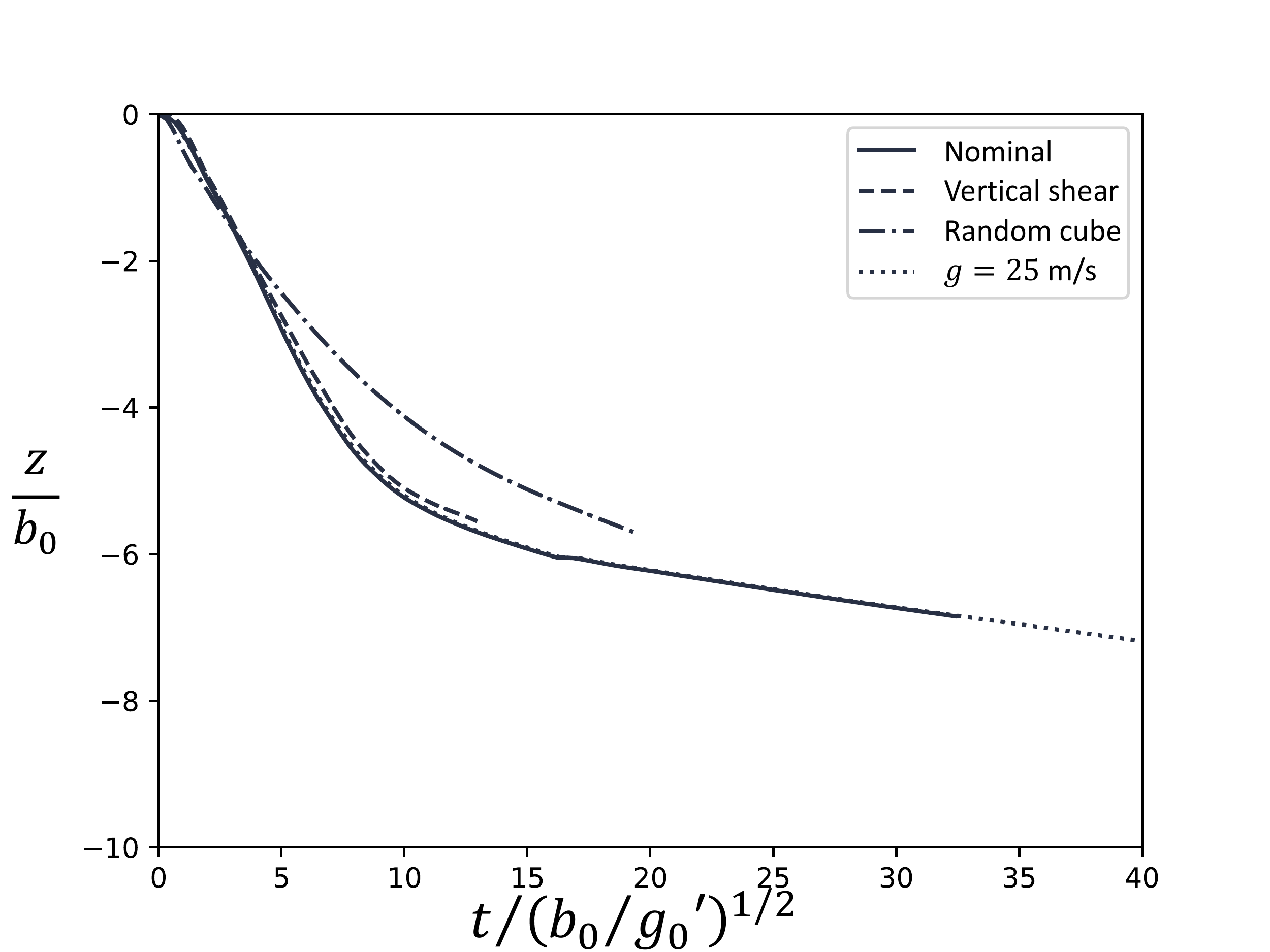}
\caption{Comparing the vertical location of the centroid for different models addressing different confounding factors: changing the gravitational field, changing the initial conditions, and introducing vertical wind shear.  It is difficult to distinguish by eye between the two assumptions for gravity, because they overlap each other almost perfectly in non-dimensionalized form.}
\label{fig:confounding}
\end{figure}

The background velocity field in our simulations is initialized to be zero.  
Besides environmental turbulence, this likewise neglects the possibility of vertical wind shear. 
On the Earth, vertical wind shear is correlated with increased storm activity. 
If we expect the same principle to apply to Jupiter and other planets, then the regions where stormy rainy downdrafts are occuring probably possess some amount of vertical shear, and we must assess the extent to which this modifies our conclusions. 
In Figure~\ref{fig:confounding} we compare zero shear to moderate vertical wind shear (wind velocity changing by 10m~s$^{-1}$ over one density scale height). 
The results are practically unaffected. 
We find that for extreme vertical wind shear (100m~s$^{-1}$ over one density scale height, not plotted), the perturbation is completely ripped apart and efficiently mixed with the environment. 
In that case, the extreme vertical wind shear produces Kelvin-Helmholtz instabilities, acoustic radiation, and vigorous dissipative turbulence such that such an extreme shear would require a large amount of energy to sustain. 
For more moderate vertical wind shear, our results are perturbed but not profoundly affected. 
On Jupiter, beneath about 4 bars (above the vaporization level for water), the vertical wind shear is very slight \citep{seiff+1997}. 
Therefore we can say with some confidence that our results are not too unrealistic for atmospheres that possess realistic vertical wind shear. 

\subsection{Stratification}
\label{subsec:stratification}
Our investigations in this work focus on a neutrally stratified environment. 
However, some observations suggest Jupiter's atmosphere may actually be weakly stably stratified \citep{allison-atkinson2001, wong+2011, guillot+2020ii, bolton+2021}. 
In this case, the thermal's acceleration will be modified by the change in potential temperature of its surroundings. 
If we specify a Br\"{u}nt-V\"{a}is\"{a}l\"{a} frequency $N$ associated with  the stratification, then Eq.~\ref{eq:buoyancy-incompressible} is modified to become \citep{turner-book}
\begin{equation}
\frac{d(b^3 g')}{dz} = - b^3 N^2.
\end{equation}
For positive real $N^2$, the solution to this equation behaves like a converging exponential decay. 
Therefore for a constant stratification gradient, there exists some depth at which the thermal will become neutrally buoyant as a combination of its dilution from entrainment and its propagation to a higher density environment. 
The depth at which this will occur can be determined using dimensional analysis
\begin{equation}
z_{\rm max} = 5 g_0'^{1/4} b_0^{3/4} N^{-1/2}.
\end{equation}
The coefficient of proportionality can be determined empirically so that this result agrees with observations over a large range of scales \citep{briggs1969}. 
If we use characteristic values for Jupiter of $b_0 \sim 25$km, $g_0' \sim 2$~m~s$^{-2}$, and $N \sim 0.02$~s$^{-1}$ \citep{wong+2011}, we obtain $z_{\rm max} \sim $~84km.  
This depth should be larger in the case of compression, because the effect of dilution from entrainment will be counteracted by the effect of compression as the thermal falls. 
Nevertheless, if Jupiter's atmosphere is really stably stratified, this could substantially affect our results. 
In general, we expect the penetration depth in a stably stratified environment to be less than for a neutrally stratified environment. \\

In abyssal atmospheres, some part of the atmosphere may be stably stratified even if it is convective or neutrally stratified above and below that layer. 
It is then natural to wonder whether a vortex ring could pierce such a layer of stability before continuing to propagate downward. 
Assuming the layer of static stability is encountered at some depth $z < z_p$ so that the thermal vortex ring structure remains intact, we can ask two types of questions. 
The first is whether the compositional gradient is extreme enough to decelerate the downwelling thermal. 
If the magnitude of $\delta \rho / \rho$ within the thermal exceeds the change in buoyancy associated with the compositional gradient. 
One way to do this is to compare the virtual potential temperature 
\begin{equation}
\theta_v \equiv \frac{T}{1 - x(1-\mu_c/\mu_d)} \left(\frac{p}{p_{\rm ref}} \right)^{-(\gamma-1)/\gamma},
\end{equation}
where $p_{\rm ref}$ is some reference pressure. 
If $\theta_v$ is the thermal is less than $\theta_v$ at the base of the stable layer, then to first approximation the thermal should pierce through to the other side. 
On Jupiter where we expect relatively weak stratification, this should occur, although in planets like Uranus with stronger stratification there may be a level of neutral buoyancy within the stable layer. 
Then one must compare the potential energy barrier from the level of neutral buoyancy to the kinetic energy within the thermal to determine whether it can pierce the stable layer. 
within the thermal and at the base of the stable layer. 
On Uranus or Neptune, this would require downdraft velocities in excess of 100~m/s \citep{markham-stevenson2021}.

\subsection{Applicability to different environments}
We conducted the majority of our tests for Figures~\ref{fig:drops}, \ref{fig:compilation}, and \ref{fig:penetration} for a particular set of environmental conditions, namely a hydrogen/helium atmosphere with an Earth-like acceleration of 9.8m~s$^{-2}$, releasing the thermal from 4 bars with a temperature of 200K. 
However, we claim that our results are robust to different assumptions about the atmosphere. 
The results are only sensitive to your choice of $g$ and $H_0$; when properly non-dimensionalized, the results will be equivalent. 
We demonstrate this in Figure~\ref{fig:confounding} by changing the gravity of the simulation, finding the non-dimensionalized simulation output to be practically indistinguishable. 
The results are the same when changing other atmospheric parameters such as the Gr\"{u}neisen parameter or molecular weight. 

\subsection{Compositional effects}

While the dynamics are robustly unaffected in a compositionally uniform atmosphere, we have not in this work addressed how compositional differences between the thermal and the environment may affect the dynamics. 
In this work we modeled density perturbations as temperature perturbations. 
This is appropriate so long as the Gr\"{u}neisen parameter is identical inside and outside the thermal, and if the condensing species is fully vaporized. 
If these assumptions are not met, however, the dynamics are affected. 
For example, ammonia and water both have more molecular rotational degrees of freedom than hydrogen and helium. 
They therefore have a smaller Gr\"{u}neisen parameter, and a less steep adiabatic gradient. 
If the environmental lapse rate is steeper than the adiabatic lapse rate within the thermal, the dynamics behave like an unstably stratified environment. 
This may further enhance the downward propagating potential of rainy downdraft laden with vapors like water, ammonia, and methane. 
The importance of this effect, however, depends on the concentration of vapor in the downdraft, something that is beyond the scope of this work. 
Additionally, microphysics may allow some fraction of the precipitation to remain condensed coexisting alongside its dense vapor phase. 
A followup work will investigate the consequences of these dynamics in greater detail, modeling the storms from their initiation through their eventual rainy downdraft phase.

\section{Conclusion and Discussion}
\label{sec:discussion}
\subsection{The penetration depth}
Through modeling and simulation, we have arrived at the robust conclusion that spatially localized strong density perturbations can propagate to $\sim 3-8$ times their initial size without substantially mixing with their environment, see Figures~\ref{fig:example},~\ref{fig:sigs}~and~\ref{fig:penetration}. 
The propagation depth depends on the ratio of the initial perturbation size to the local density scale height, $b_0/H_0$. 
We define the penetration depth either according to the vertical spreading of material increasing by a factor of $e$ above its original distribution, or according to velocity flattening where the centroid subsidence velocity falls below some defined value. 
We find good agreement between either criterion. 
However, both criteria are arbitrary; really in neither case does subsidence stop entirely. 
The material continues to subside beyond the point at which detailed simulations become computationally impractical. 
Eventually unmodeled effects, such as environmental turbulence or stratification, is needed to halt the subsidence---but there is nothing fundamental preventing a buoyancy anomaly from subsiding indefinitely. 
That being said, the downdraft remains quantifiably coherent and concentrated down to significant depth without significantly mixing with its environment. 
For thermals of order the size of a density scale height, they can propagate about three density scale heights down without mixing. 
For much smaller perturbations ten times smaller than a density scale height, they can still propagate coherently down to eight times their initial size, or almost a full density scale height. \\

\subsection{Depletion of volatiles in giant planet atmospheres}
Applied to Jupiter, this lends credence to the notion suggested in \cite{guillot+2020ii} that even after the vaporization of volatile species, downdrafts can continue to sink coherently without efficiently mixing with their environment. 
Our results also suggest that the compositional gradients observed in the volatile abundance on Jupiter may exist on other planets, even without invoking the ammonia-water phase mixture relevant on Jupiter. 
A rainy downdraft that vaporizes quickly upon reaching its boiling point as it descends will still represent a spatially localized density perturbation. 
The subsequent dynamics would play out along the lines of our simulated results, with the perturbation maintaining some coherence over relatively long length scales; on Jupiter, of order a hundred kilometers. \\

Our results indicate that a pond of rain originating from a source on this scale should be expected to propagate down to tens of bars before efficiently mixing with its environment. 
It will be of interest to assess the particularity or specificity of the compositional gradients in Jupiter's volatile distribution observed by Juno. 
Here we present a plausible mechanism by which a planet with a simpler hydrological cycle could exhibit similar behavior. 
Uranus, for example, exhibits energetic methane storms \citep{depater+2015}. 
Rain from such events could plausibly penetrate to great depth before mixing with its surroundings. 
Measuring whether Uranus' deep methane abundance beneath the clouds is homogenous or variable will be of great interest for an upcoming mission. 
Comparative planetology between Uranus and Jupiter will be important for interpretting spectra from exoplanets, which may be depleted in volatiles in the observable atmospheres if they are stormy. 
These results are also of interest to the question of the depth of the weather layer on giant planets. 
Our results suggest that the weather layer associated with a condensing species should extend substantially below its cloud layer. 
In the context of Jupiter, our results suggest the weather layer should extend at least to tens of bars. \\

\subsection{On non-local convection in the Sun}
We now compare our results and conclusions to prior work that has investigated similar phenomena. 
In particular, \cite{anders+2019} investigate the concept of a localized thermal in a highly compressible environment in the context of the Sun, motivated by the notion of entropy rain. 
In that work, the authors found results in agreement with our predictions from Sect.~\ref{sec:turner}, but did not produce our observed braking instability from Sect.~\ref{sec:results}. 
The primary cause for the differences between our findings was our different treatment of turbulence; their work modeled the laminar case, while this work considers the turbulent case.  
We discussed these differences in more detail in Section~\ref{sec:viscosity}~and~\ref{sec:applications_sun}. 
Concerning entropy rain as described by \cite{brandenburg2016}, our results do not directly corroborate the \cite{anders+2019} finding that the phenomenon can be a simple consequence of atmospheric compression. 
However, our findings do not rule out the possibility of non-local convection if coherence can be enforced by some other mechanism, for example by considering rotation or magnetic field effects.
\\

\subsection{Outstanding problems and future work}
Here we focused particularly on tracking the motion of vaporized rain. 
Future steps for a more complete understanding will involve modeling the complete hydrological cycle, including upwelling columns and homogeous turbulent convection. 
As discussed in Sect.~\ref{sec:confounding}, our results are impacted by more complex but realistic planetary environments. 
Layers of stable stratification, likely to exist at least intermittently as they do on Earth, as well as environmental turbulence and vertical wind shear likely have an effect on the output. 
Additionally, the initial conditions we chose for this study were largely ad hoc; a fully self-consistent model including storm development and precipitation microphysics would be a reasonable next step. 
Here we have found, however, that simply invoking a mixing length on the order a scale height for the downdraft coherence lengthscale may actually be an underestimate. 
Indeed, downdrafts can maintain coherence without mixing with their environment down to multiple scale heights, in the context of Jupiter on the order of a hundred kilometers or tens of bars. 
Planets are natural laboratories of exotic physical processes. 
We must visit more to learn more about the poorly understood physics, and the range of possible planetary climates.

\section*{Data availability}
Sample simulation animations for visualization and intuition available at \href{https://doi.org/10.5281/zenodo.7705225}{10.5281/zenodo.7705225}. 
Hydrodynamic output data in netCDF format and accompanying analysis software available upon request. 

\section*{Acknowledgements}
This work has been funded by the CNES postdoctoral fellowship program. 
Computations were performed using the Licallo supercomputing cluster hosted by Observatoire de la C\^{o}te d'Azur. 
We would like to thank Jeremie Bec and Holger Homann for instructive discussions about resolution dependence and closure models for turbulence.
We would also like to thank the anonymous reviewer for constructive feedback.

\bibliography{library}{}

\begin{thebibliography}{}

\bibitem[Allison and Atkinson, 2001]{allison-atkinson2001}
Allison, M. and Atkinson, D. (2001).
\newblock Galileo probe doppler residuals as the wave-dynamical signature of a
  weakly stable, downard-increasing stratification in jupiter's deep wind
  layer.
\newblock {\em GRL}, 28(14):2747--2750.

\bibitem[Anders et~al., 2019]{anders+2019}
Anders, E., Lecoanet, D., and Brown, B. (2019).
\newblock Entropy rain: Dilution and compression of thermals in stratified
  domains.
\newblock {\em ApJ}, 884(65).

\bibitem[Becker et~al., 2020]{becker+2020}
Becker, H., Alexander, J., Atreya, S., Bolton, S., Brennan, M., et~al. (2020).
\newblock Small lightning flashes from shallow electrical storms on jupiter.
\newblock {\em Nature}, 584:55--58.

\bibitem[Bjoraker et~al., 2015]{bjoraker+2015}
Bjoraker, G., Wong, M., de~Pater, I., and \'{A}d\'{a}kovics, M. (2015).
\newblock Jupiter's deep cloud structure revealed using keck observations of
  spectrally resolved line shapes.
\newblock {\em ApJ}, 810(2).

\bibitem[Bjoraker et~al., 2018]{bjoraker+2018}
Bjoraker, G., Wong, M., de~Pater, I., Hewagama, T., et~al. (2018).
\newblock The gas composition and deep cloud structure of jupiter's great red
  spot.
\newblock {\em AJ}, 156(101).

\bibitem[Bolton et~al., 2021]{bolton+2021}
Bolton, S., Levin, S., Guillot, T., Li, C., Kaspi, Y., et~al. (2021).
\newblock Microwave observations reveal the deep extent and structure of
  jupiter's atmospheric vortices.
\newblock {\em Science}, 374(6570):968--972.

\bibitem[Borucki et~al., 1982]{borucki+1982}
Borucki, W., Bar-Nun, A., Scarf, F., Cook, A., and Hunt, G. (1982).
\newblock Lightning activity on jupiter.
\newblock {\em Icarus}, 52(3):492--502.

\bibitem[Brandenburg, 2016]{brandenburg2016}
Brandenburg, A. (2016).
\newblock Stellar mixing length theory with entropy rain.
\newblock 832(6).

\bibitem[Briggs, 1969]{briggs1969}
Briggs, G. (1969).
\newblock Iii. mathematical analysis of chimney plume rise and dispersion.
\newblock {\em Phil. Trans. Roy. Soc. Lond. A}, 265:197--203.

\bibitem[Brown et~al., 2018]{brown+2018}
Brown, S., Janssen, M., Adumitroaie, V., Atreya, S., Bolton, S., et~al. (2018).
\newblock Prevalent llightning sferics at 600 megahertz near jupiter's poles.
\newblock {\em Nature}, 558:87--90.

\bibitem[Brummell et~al., 2002]{brummell+2002}
Brummell, N., Clune, T., and Toomre, J. (2002).
\newblock Penetration and overshooting in turbulent compressible convection.
\newblock {\em ApJ}, 570:825--854.

\bibitem[Cattaneo et~al., 2003]{cattaneo+2003}
Cattaneo, F., Emonet, T., and Weiss, N. (2003).
\newblock On the interaction between convection and magnetic fields.
\newblock {\em ApJ}, 588(2).

\bibitem[Chang and Sofia, 1987]{chan-sofia1987}
Chang, K. and Sofia, S. (1987).
\newblock Validity tests of the mixing-length theory of deep convection.
\newblock {\em Science}, 235(4787):465--467.

\bibitem[Conrath et~al., 1991]{conrath+1991}
Conrath, B., Flasar, F., and Gierasch, P. (1991).
\newblock Thermal structure and dynamics of neptune's atmosphere from voyager
  measurements.
\newblock {\em JGR Space Phys.}, 96(S01):18931--18939.

\bibitem[de~Pater et~al., 2019]{depater+2019}
de~Pater, I., Sault, R., Wong, M., Fletcher, L., DeBoer, D., and Butler, B.
  (2019).
\newblock Jupiter's ammonia distribution derived from vla maps at 3-37 ghz.
\newblock {\em Icarus}, 322:168--191.

\bibitem[de~Pater et~al., 2015]{depater+2015}
de~Pater, I., Sromovsky, L., Fry, P., Hammel, H., Baranec, C., and Sayanagi, K.
  (2015).
\newblock Record-breaking storm activity on uranus in 2014.
\newblock {\em Icarus}, 252:121--128.

\bibitem[Del~Genio and McGrattan, 1990]{delgenio-mcgrattan1990}
Del~Genio, A. and McGrattan, K. (1990).
\newblock Moist convection and the vertical structure and water abundance of
  jupiter's atmosphere.
\newblock {\em Icarus}, 84(1):29--53.

\bibitem[Fischer et~al., 2011]{fischer+2011}
Fischer, G., Kurth, W., Gurnett, D., Zarka, P., Kyudina, U., and other (2011).
\newblock A giant thunderstorm on saturn.
\newblock {\em Nature}, 475:75--77.

\bibitem[Gao and Yu, 2016]{gao-yu2016}
Gao, L. and Yu, S. (2016).
\newblock Vortex ring formation in starting forced plumes with negative and
  positive buoyancy.
\newblock {\em Phys. Fluids}, 28(113601).

\bibitem[Gibbard et~al., 1999]{gibbard+1999}
Gibbard, S., Levy, E., Lunine, J., and de~Pater, I. (1999).
\newblock Lightning on neptune.
\newblock {\em Icarus}, 139(2):227--234.

\bibitem[Gierasch et~al., 2000]{gierasch+2000}
Gierasch, P., Ingersoll, A., Banfield, D., Ewald, S.P. nd~Helfenstein, P.,
  et~al. (2000).
\newblock Observation of moist convection in jupiter's atmosphere.
\newblock {\em Nature}, 403:628--630.

\bibitem[Guillot, 1995]{guillot1995}
Guillot, T. (1995).
\newblock Condensation of methane, ammonia, and water and the inhibition of
  convection in giant planets.
\newblock {\em Science}, 269(5231):1697--1699.

\bibitem[Guillot et~al., 2020a]{guillot+2020ii}
Guillot, T., Li, C., Bolton, S., Brown, S., Ingersoll, A., et~al. (2020a).
\newblock Storms and the depletion of ammonia in jupiter: Ii. explaining the
  juno observations.
\newblock {\em JGR Planets}, 125(8).

\bibitem[Guillot et~al., 2020b]{guillot+2020i}
Guillot, T., Stevenson, D., Atreya, S., Bolton, S., and Becker, H. (2020b).
\newblock Storms and the depletion of ammonia in jupiter: I. microphysics of
  ``mushballs''.
\newblock {\em JGR Planets}, 125(8).

\bibitem[Guillot et~al., 2004]{guillot+2004}
Guillot, T., Stevenson, D., Hubbard, W., and Didier, S. (2004).
\newblock The interior of jupiter.
\newblock In Bagenal, F., Dowling, T., and McKinnon, W., editors, {\em Jupiter,
  The planet, satellites and magnetosphere.}, chapter~4, pages 35--57.
  Cambridge University Press, Cambridge, UK.

\bibitem[Hanasoge et~al., 2012]{hanasoge+2012}
Hanasoge, S., Duvall, T., and Sreenivasan, K. (2012).
\newblock Anomalously weak solar convection.
\newblock {\em PNAS}, 109(30):11928--11932.

\bibitem[Heath and McKim, 1992]{heath-mckim1990}
Heath, A. and McKim, R. (1992).
\newblock Saturn 1990: The great white spot.
\newblock {\em J. Br. Astronom. Assoc.}, 102(4).

\bibitem[Helling, 2019]{helling2019}
Helling, C. (2019).
\newblock Exoplanet clouds.
\newblock {\em Ann. Rev. EPS}, 47:583--606.

\bibitem[Helling and Casewell, 2014]{helling-casewell2014}
Helling, C. and Casewell, S. (2014).
\newblock Atmospheres of brown dwarfs.
\newblock {\em A\&:A Rev.}, 22(80).

\bibitem[Hueso et~al., 2002]{hueso+2002}
Hueso, R., S\'{a}nchez-Lavega, A., and Guillot, T. (2002).
\newblock A model for large-scale convective storms in jupiter.
\newblock {\em JGR Planets}, 107(E10):5--11.

\bibitem[Hurlburt et~al., 1984]{hurlburt+1984}
Hurlburt, N., Toomre, J., and Massaguer, J. (1984).
\newblock Two-dimension compressible convection extending over multiple scale
  heights.
\newblock {\em ApJ}, 282:557--573.

\bibitem[Kolmasova et~al., 2018]{kolmasova+2018}
Kolmasova, I., Imai, M., Satolik, O., Kurth, W., Hospardarsky, G., et~al.
  (2018).
\newblock Discovery of rapid whistlers close to jupiter implying lightning
  rates similar to those on earth.
\newblock {\em Nature Astronomy}, 2:544--548.

\bibitem[Launder and Spalding, 1972]{launder-spalding1972}
Launder, B. and Spalding, D. (1972).
\newblock Mathematical models of turbulence.

\bibitem[Li and Chen, 2019]{snap}
Li, C. and Chen, X. (2019).
\newblock Simulating nonhydrostatic atmospheres on planets (snap): Formulation,
  validation, and application to the jovian atmosphere.
\newblock {\em ApJS}, 240(2).

\bibitem[Li and Ingersoll, 2015]{li-ingersoll2015}
Li, C. and Ingersoll, A. (2015).
\newblock Moist convection in hydrogen atmospheres and the frequency of
  saturn's giant storms.
\newblock {\em Nature Geoscience}, 8:398--403.

\bibitem[Li et~al., 2020]{li+2020}
Li, C., Ingersoll, A., Bolton, S., Levin, S., Janssen, M., et~al. (2020).
\newblock The water abundance in jupiter's equatorial zone.
\newblock {\em Nature Astronomy}, 4:609--616.

\bibitem[Li et~al., 2017]{li+2017}
Li, C., Ingersoll, A., Janssen, M., Levin, S., Bolton, S., et~al. (2017).
\newblock The distribution of ammonia in jupiter from a preliminary inversion
  of juno microwave radiometer data.
\newblock {\em GRL}, 44(11):5317--5325.

\bibitem[Lindal et~al., 1987]{lindal+1987}
Lindal, G., Lyons, J., Sweetnam, D., Eshleman, V., Hinson, D., and Tyler, G.
  (1987).
\newblock The atmosphere of uranus: Results of radio occultation measurements
  with voyager 2.
\newblock {\em JGR Space Physics}, 92(A13).

\bibitem[Little et~al., 1999]{little+1999}
Little, B., Anger, C., Ingersoll, A., Vasavada, A., Sneske, D., et~al. (1999).
\newblock Galileo images of lightning on jupiter.
\newblock {\em Icarus}, 142(2):306--323.

\bibitem[Loftus et~al., 2019]{loftus+2019}
Loftus, K., Wordsworth, R., and Morley, C. (2019).
\newblock Sulfate aerosol hazes and so2 gas as constraints on rocky exoplanets'
  surface liquid water.
\newblock {\em ApJ}, 887(231).

\bibitem[Markham and Stevenson, 2018]{markham-stevenson2018}
Markham, S. and Stevenson, D. (2018).
\newblock Excitation mechanisms for jovian seismic modes.
\newblock {\em Icarus}, 306:200--213.

\bibitem[Markham and Stevenson, 2021]{markham-stevenson2021}
Markham, S. and Stevenson, D. (2021).
\newblock Constraining the effect of convective inhibition on the thermal
  evolution of uranus and neptune.
\newblock {\em PSJ}, 2(146).

\bibitem[Molter et~al., 2019]{molter+2017}
Molter, E., de~Pater, I., Luszcz-Cook, S., Hueso, R., Tollefson, J., et~al.
  (2019).
\newblock Analysis of neptune's 2017 bright equatorial storm.
\newblock {\em Icarus}, 321:324--345.

\bibitem[Molter et~al., 2021]{molter+2021}
Molter, E., de~Pater, I., Luszcz-Cook, S., Tollefson, J., et~al. (2021).
\newblock Tropospheric composition and circulation of uranus with alma and the
  vla.
\newblock {\em PSJ}, 2(1).

\bibitem[Morton et~al., 1956]{morton+1956}
Morton, B., Taylor, G., and Turner, J. (1956).
\newblock Turbulent gravitational convection from maintained and instantaneous
  sources.
\newblock {\em Roy. Soc.}, 234(1196).

\bibitem[Nordlund et~al., 2009]{nordlund+2009}
Nordlund, A., Stein, R., and Asplund, M. (2009).
\newblock Solar surface convection.
\newblock {\em Living Rev. Solar Phys.}, 6(2).

\bibitem[Pope, 2000]{pope2000}
Pope, S. (2000).
\newblock {\em Turbulent Flows}.
\newblock Cambridge University Press, Cambridge.

\bibitem[Roe, 1981]{roe1981}
Roe, P. (1981).
\newblock Approximate riemann solvers, parameter vectors, and difference
  schemes.
\newblock {\em J. Comp. Phys.}, 43(2):357--372.

\bibitem[Seiff et~al., 1997]{seiff+1997}
Seiff, A., Blanchard, R., Knight, T., Schuber, G., Kirk, D., et~al. (1997).
\newblock Wind speeds measured in the deep jovian atmosphere by the galileo
  probe acceleratometers.
\newblock {\em Nature}, 388:650--652.

\bibitem[Shusser and Gharib, 2000]{shusser-gharib2000}
Shusser, M. and Gharib, M. (2000).
\newblock A model for vortex ring formation in a starting buoyant plume.
\newblock {\em JFM}, 416:173--185.

\bibitem[Siebesma et~al., 2007]{siebesma+2007}
Siebesma, A., Soares, P., and Teixeira, J. (2007).
\newblock A combined eddy-diffusivity mass-flux approach for the convective
  boundary layer.
\newblock {\em JAtS}, 64(4):1230--1248.

\bibitem[Sromovsky et~al., 2011]{sromovsky+2011}
Sromovsky, L., Fry, P., and Kim, J. (2011).
\newblock Methane on uranus: The case for a compact ch4 cloud layer at low
  latitutdes and a severe ch4 depletion at high-latitudes based on re-analysis
  of voyager occultation measurements and stis spectroscopy.
\newblock {\em Icarus}, 215(1):293--312.

\bibitem[Straka et~al., 1993]{straka+1993}
Straka, J., Wilhelmson, R., Wicker, L., Anderson, J., and Droegemeier, K.
  (1993).
\newblock Numerical solutions of a non-linear density current: A benchmark
  solution and comparisons.
\newblock {\em Num. Meth. Fluids}, 17(1):1--22.

\bibitem[Suselj et~al., 2019]{suselj+2019}
Suselj, K., Kurowski, M., and Teixeira, J. (2019).
\newblock A unified eddy-diffusivity/mass-flux approach for modeling
  atmospheric convection.
\newblock {\em JAtS}, 76(8):2505--2537.

\bibitem[Tollefson et~al., 2021]{tollefson+2021}
Tollefson, J., de~Pater, I., Molter, E., Sault, R., Bulter, B., et~al. (2021).
\newblock Neptune's spatial brightness temperature variations from the vla and
  alma.
\newblock {\em PSJ}, 2(3).

\bibitem[Toro, 2013]{toro2013}
Toro, E. (2013).
\newblock {\em Riemann Solvers and Numerical Methods for Fluid Dynamics}.
\newblock Springer, Heidelberg.

\bibitem[Turner, 1957]{turner1957}
Turner, J. (1957).
\newblock Buoyant vortex rings.
\newblock {\em Proc. Roy. Soc. A}, 239(1216).

\bibitem[Turner, 1962]{turner1962}
Turner, J. (1962).
\newblock The motion of buoyant elements in turbulent surroundings.
\newblock {\em Fluid Mech.}, 16(1).

\bibitem[Turner, 1963]{turner1963}
Turner, J. (1963).
\newblock The flow into an expanding spherical vortex.
\newblock {\em JFM}, 18(2).

\bibitem[Turner, 1973]{turner-book}
Turner, J. (1973).
\newblock {\em Buoyancy Effects in Fluids}.
\newblock Cambridge University Press, Cambridge.

\bibitem[Wang and Geerts, 2013]{wang-geerts2013}
Wang, Y. and Geerts, B. (2013).
\newblock Composite vertical structure of vertical velocity in nonprecipitating
  cumulus clouds.
\newblock {\em Mon. Weather Rev.}, 141(5):1673--1692.

\bibitem[Weiss et~al., 2004]{weiss+2004}
Weiss, N., Thomas, J., Brummell, N., and Tobias, S. (2004).
\newblock The origin of penumbral structure in sunspots: Downward pumping of
  magnetic flux.
\newblock {\em ApJ}, 600(2).

\bibitem[Wong et~al., 2011]{wong+2011}
Wong, M., de~Pater, I., Asay-Davis, X., Marcus, P., and Go, C. (2011).
\newblock Vertical structure of jupiter's oval ba before and after it reddened:
  What changed?
\newblock {\em Icarus}, 215(1):211--225.

\bibitem[Wong et~al., 2004]{wong+2004}
Wong, M., Mahaffy, P., Atreya, S., Niemann, H., and Owen, T. (2004).
\newblock Updated galileo probe mass spectrometer measurements of carbon,
  oxygen, nitrogen, and sulfur on jupiter.
\newblock {\em Icarus}, 171(1):153--170.

\bibitem[Zhou et~al., 2020]{zhou+2020}
Zhou, X., Xu, Y., and Zhang, W. (2020).
\newblock Formation regimes of vortex rings in thermals.
\newblock {\em J. Fluid Mech.}, 885(A44).

\bibitem[Zhu and Dong, 2021]{zhu-dong2021}
Zhu, W. and Dong, S. (2021).
\newblock Exoplanet statistics and theoretical implications.
\newblock {\em Ann. Rev. A\&:A}, 59:291--336.

\end{thebibliography}
\bibliographystyle{apalike}

\end{document}